\documentclass[a4paper,11pt]{article}

\pdfoutput=1 
\usepackage{./OnShellActionFiles/jheppub}
\usepackage[T1]{fontenc}
\usepackage{lmodern}

\usepackage[T1]{fontenc}
\usepackage[utf8]{inputenc}
\usepackage[english]{babel}
\usepackage{graphicx}
\usepackage{amssymb}
\usepackage{mathrsfs}
\usepackage{microtype}
\usepackage{booktabs}
\usepackage{array, tabularx}
\usepackage{chngpage}
\usepackage{braket}
\usepackage{bbm}
\usepackage{mathtools}
\usepackage{amsthm}
\usepackage[english]{varioref}
\usepackage{url}
\usepackage{multicol, multirow}
\usepackage{mparhack}
\usepackage{emptypage}
\usepackage{setspace}
\usepackage{dsfont}
\usepackage{rotating}
\usepackage{enumitem}
\usepackage{comment}
\usepackage{tikz,tikz-cd}
\usepackage{xcolor}

\usetikzlibrary{cd,shadings}

\numberwithin{equation}{section}  

\newcommand{\beq}{\begin{equation}}
\newcommand{\eeq}{\end{equation}}
\newcommand{\nn}{\nonumber}

\newcommand{\numberset}{\mathbb}

\newcommand{\R}{\numberset{R}}
\newcommand{\C}{\numberset{C}}
\newcommand{\Z}{\numberset{Z}}

\newcommand{\mb}[1]{\mathbf{#1}}
\newcommand{\mc}[1]{\mathcal{#1}}

\newcommand{\cC}{\mc{C}}

\newcommand{\cL}{\mc{L}}
\newcommand{\cN}{\mc{N}}
\newcommand{\cO}{\mc{O}}

\newcommand{\cV}{\mc{V}}

\newcommand{\rd}{{\rm d}}

\newcommand{\ii}{{\rm i}}
\newcommand{\identity}{\mathds{1}}

\newcommand{\e}{{\rm e}}
\newcommand{\E}{{\rm E}}
\newcommand{\vol}{{\rm vol}}
\newcommand{\hook}{\mathbin{\rule[.2ex]{.4em}{.03em}\rule[.2ex]{.03em}{.9ex}}}
\newcommand{\zbar}{\overline{z}}

\newcommand{\Ric}{{\rm Ric}}

\renewcommand{\Re}{\text{Re}}
\renewcommand{\Im}{\text{Im}}

\renewcommand{\j}{\varphi}
\newcommand{\empot}{\nu}
\newcommand{\nutpot}{\sigma}

\DeclarePairedDelimiter{\abs}{\lvert}{\rvert}

\DeclareMathOperator{\arcsinh}{arcsinh}
\DeclareMathOperator{\sgn}{sgn}

\allowdisplaybreaks 

\newcommand{\nocontentsline}[3]{}
\newcommand{\tocless}[2]{\bgroup\let\addcontentsline=\nocontentsline#1{#2}\egroup}
\title{Localization of the action in AdS/CFT}

\author[a]{Pietro Benetti Genolini,}
\author[b]{Juan Manuel P\'erez Ipi\~na,}
\author[b]{and James Sparks}

\affiliation[a]{Department of Applied Mathematics and Theoretical Physics, University of Cambridge, Wilberforce Road, Cambridge, CB3 OWA, UK}
\affiliation[b]{Mathematical Institute, University of Oxford, Woodstock Road, Oxford, OX2 6GG, UK}

\emailAdd{Pietro.BenettiGenolini@damtp.cam.ac.uk}
\emailAdd{Juan.PerezIpina@maths.ox.ac.uk}
\emailAdd{James.Sparks@maths.ox.ac.uk}

\abstract{
We derive a simple formula for the action of any supersymmetric solution to minimal gauged supergravity in the AdS$_4$/CFT$_3$ correspondence. 
Such solutions are equipped with a supersymmetric Killing vector, and we show that the holographically renormalized action may be expressed 
entirely in terms of the weights of this vector field at its fixed points, together with certain topological data. In this sense, the 
classical gravitational partition function localizes in the bulk. We illustrate our general formula 
with a number of explicit examples, in which exact dual field theory computations are also available, which include supersymmetric Taub-NUT and Taub-bolt type 
spacetimes, as well as black hole solutions. Our simple topological formula also allows us to write down the action of any solution, provided it exists.}

\begin{document} 

\maketitle

\section{Introduction and summary}
\label{sec:Introduction}

Localization has played a key role in recent developments in AdS/CFT, allowing for the exact computation of certain field theory observables at strong coupling \cite{Pestun:2016zxk}. These computations may in turn be compared to classical, or semi-classical, gravity. Here the localization refers to supersymmetry in the field theory path integral, where under favourable circumstances the latter
reduces to a well-defined finite-dimensional integral. In this paper we show that localization also plays a role in the dual classical gravity computation: specifically, 
we show that the holographically renormalized action $I$ of any Euclidean supersymmetric solution to minimal gauged supergravity in four dimensions 
localizes to the fixed points of a supersymmetric Killing vector. Since $\exp(-I)$ is identified with the gravitational partition function, in a saddle point approximation, 
in this sense the classical gravitational partition function localizes in the bulk. This should be distinguished from recent work attempting to define localization in a supergravity path integral -- 
see, for example, \cite{Dabholkar:2014wpa, Hristov:2018lod, deWit:2018dix, Jeon:2018kec}.\footnote{Of course, it is tempting to speculate that localization in both senses 
plays a role in gravity, as in the seminal field theory work of \cite{Nekrasov:2002qd}.}

One of the motivations for the present work was to try to understand what are the fundamental structures underpinning certain supersymmetric observables in AdS/CFT. 
Rather than regard the agreement of  dual computations of a particular observable, in a particular theory, as ``miraculous'', one would like to go further and try 
to understand at a more fundamental level why the computations are equivalent. To this end, we here focus on solutions to minimal $\mathcal{N}=2$ gauged supergravity in four dimensions \cite{Freedman:1976aw}. The bosonic sector of this theory is Einstein--Maxwell theory with a negative cosmological constant. There are a number of reasons for focusing on this case. Two points worth mentioning here are that, firstly, solutions may be uplifted on different internal 7-manifolds $Y_7$ to obtain solutions of M-theory \cite{Gauntlett:2007ma}, 
with known dual three-dimensional supersymmetric gauge theories; and secondly, this supergravity theory is simple, with a number of explicitly known solutions 
whose gravitational partition functions have already been matched to strong coupling (large $N$) exact field theory partition functions \cite{Martelli:2011fu, Martelli:2011fw, Martelli:2012sz, Martelli:2013aqa, Farquet:2014kma, Azzurli:2017kxo, Toldo:2017qsh}. 

In the remainder of this introduction we summarize our main result. As already mentioned, we consider
Euclidean supersymmetric solutions to minimal $\mathcal{N}=2$ gauged supergravity  that are asymptotically locally AdS. We denote the bulk 
4-manifold as $M$, with the conformal boundary 3-manifold $M_3=\partial M$. 
Every supersymmetric solution is equipped with a canonical Killing vector field $\xi$ on $M$, defined as a bilinear in the Killing spinor. We take this to be nowhere zero on
 the boundary $M_3$, so that the induced rigid supersymmetric geometry on $M_3$ is that in \cite{Closset:2012ru}.\footnote{This is not strictly necessary for the gravity computation, but it \emph{is} necessary to compare 
with field theory localization results that have been developed to date.} As usual, 
the on-shell action $I$ for such a solution is divergent, but may be regularized by adding boundary counterterms \cite{Emparan:1999pm, Skenderis:2002wp}. We show that
this holographically renormalized action may be written as
\beq\label{Iintro}
I  = \left(\sum_{\mathrm{nuts}_\mp} \pm \frac{(b_1\pm b_2)^2}{4b_1b_2} + \sum_{\mathrm{bolts}\ \Sigma_\pm} \int_{\Sigma_\pm} \left(\tfrac{1}{2}c_1(T\Sigma_\pm) \mp \tfrac{1}{4}c_1(N\Sigma_\pm)\right) \right)\frac{\pi}{2G_4}~.
\eeq
The overall factor of $\pi/2G_4$, where $G_4$ denotes the four-dimensional Newton constant, is simply the action of Euclidean AdS$_4$.
The terminology in the  sums is taken from \cite{Gibbons:1979xm}: the fixed point set of $\xi$ lies in the interior of $M$, with connected components being either fixed points, called \emph{nuts}, 
or fixed two-dimensional submanifolds, called \emph{bolts}. As we show later in the paper, at such a zero of $\xi$ the bulk Dirac Killing spinor necessarily becomes chiral, and the $\pm$ 
signs in the sums over nuts$_\pm\in M$ and bolts $\Sigma_\pm\subset M$ in (\ref{Iintro}) denote chirality. At a nut isolated fixed point we may write the vector field $\xi$ as
\beq
\xi = b_1\partial_{\varphi_1}+ b_2\partial_{\varphi_2}~.
\eeq
Here the tangent space at the fixed point is $\R^4=\R^2\oplus\R^2$, with $\varphi_1$, $\varphi_2$ being standard polar angle coordinates on each copy of $\R^2\cong \C$, respectively. 
As we also explain later, due to the lack of a canonical choice of certain signs, and indeed any canonical normalization for $\xi$, 
in fact 
only the ratio $b_1/b_2\in \R\setminus \{0\}$ is well-defined. The action (\ref{Iintro}) is then a function of this ratio. 
On the other hand, a bolt is a fixed closed two-manifold $\Sigma_\pm\subset M$, with the $\pm$ signs again denoting chirality of the spinor over each connected component. Here 
$\int_{\Sigma_\pm}c_1(T\Sigma_\pm)=2-2g$ is the Chern number of the tangent bundle of the Riemann surface $\Sigma_\pm$, of genus $g$, while 
$c_1(N\Sigma_\pm)$ denotes the first Chern class of the normal bundle of $\Sigma_\pm$ in $M$. 
The bolt contribution in (\ref{Iintro}) is then a topological invariant.

Using (\ref{Iintro}) we may 
immediately reproduce all known results in the literature. Various families of explicit solutions on $M\cong \R^4$, with $\partial M=M_3\cong S^3$, 
were constructed in \cite{Martelli:2011fu, Martelli:2011fw, Martelli:2013aqa}, and in \cite{Farquet:2014kma} the formula (\ref{Iintro}) 
was proven for a general class of such ``self-dual'' solutions with this topology, where  there is a single 
isolated fixed point at the origin of $M\cong \R^4$. On the other hand, in \cite{Martelli:2012sz, Toldo:2017qsh} a class of 1/4 BPS 
``bolt'' solutions was constructed. There are two branches of solutions, referred to as bolt$_\mp$, where $M\cong 
\cO(-p)\rightarrow \Sigma_g$, so that $\partial M=M_3$ is the total space of a degree $p$ circle bundle over a Riemann surface $\Sigma_g$ of genus $g$. 
Here the supersymmetric Killing vector $\xi$ fixes the zero section $\Sigma_\pm=\Sigma_g\subset M$, and 
we may immediately read off $\int_{\Sigma_\pm} c_1(N\Sigma_\pm)=-p$ as the degree of the line bundle $\cO(-p)$. 
The formula (\ref{Iintro}) then agrees with
the results of \cite{Martelli:2012sz, Toldo:2017qsh}, taking upper and lower signs for bolt$_\mp$, respectively, where in \cite{Toldo:2017qsh} this was also obtained from a dual field theory computation. 
When $p=0$ we have $M\cong \R^2 \times \Sigma_g$, and the action~(\ref{Iintro}) reproduces minus the entropy of the black hole solution in \cite{Azzurli:2017kxo}, related to the so-called 
universal twist (see section \ref{subsec:BH}), which again has been reproduced in field theory \cite{Azzurli:2017kxo}. We also show that (\ref{Iintro}) correctly reproduces 
the action of the 1/2 BPS family of ``bolt$_\pm$'' solutions in \cite{Martelli:2012sz}, with topology $M\cong \cO(-p)\rightarrow S^2$, where in the terminology of 
the present paper the supersymmetric Killing vector has two isolated fixed points at the north and south poles of the zero section $S^2\subset M$, 
which are a nut$_\pm$ and a nut$_\mp$, respectively.  

In addition to reproducing all known results, the formula \eqref{Iintro} may also be used to compute the action of solutions \emph{assuming they exist}. 
Without a general existence theorem for supersymmetric solutions to minimal $\mathcal{N}=2$ gauged supergravity, this sort of application is 
at present somewhat formal. However, we show that the right hand side of \eqref{Iintro} may be computed very readily
for ``toric'' four-manifolds $M$, which by definition admit a $T^2$ action that contains the isometry generated by the supersymmetric Killing vector $\xi$. 
This includes all of the explicit examples mentioned in the previous paragraph (with genus $g=0$, when relevant) as special cases, but significantly generalizes them.
In particular, we consider 
 fillings of Lens spaces $L(p,q)$, and discuss the general behaviour of \eqref{Iintro} under a blow-up. 

We conclude this summary by noting that the form of (\ref{Iintro}) is very suggestive that there is a localization/fixed point theorem 
directly underlying this formula. The action depends only on topological fixed point data: the weights $b_1,b_2$ of the supersymmetric 
Killing vector $\xi$ at its isolated fixed points, Chern numbers at the fixed submanifolds, and where certain signs are determined by chirality data at the fixed points. Its form 
very much resembles an equivariant index theorem, or the Berline--Vergne/Duistermaat--Heckman fixed point theorems. 
On the other hand, the localization here is in the bulk of a 4-manifold $M$ that fills the boundary 3-manifold 
$M_3=\partial M$ on which the dual field theory is defined.  AdS/CFT states that $\exp(-I)$ should agree with the strong coupling (large $N$) partition function 
of this theory on $M_3$, where more precisely one should take the bulk solution of least action. We shall return to some of these comments 
in the discussion section  \ref{sec:Conclusions}.

\subsection*{Outline}

In section \ref{sec:4dSUGRA} we introduce the four-dimensional supergravity theory of interest, and outline the process of reduction to a three-dimensional base space for solutions with at least a $U(1)$ isometry. The resulting equations of motion can be used to show that the on-shell action is naturally exact, as in \cite{Gibbons:1979xm}. All supersymmetric solutions possess a Killing vector, and in section \ref{sec:EvalOSAction} we evaluate the on-shell action in terms of the geometric data describing the Killing vector near the fixed points of the isometry. A number of explicit examples of supersymmetric solutions are presented in section \ref{sec:Examples}, and we conclude in section \ref{sec:Conclusions} with some discussion of future directions, and possible extensions of our work.

\section{Four-dimensional gauged supergravity and reduction}
\label{sec:4dSUGRA}

\subsection{Action and equations of motion}
\label{subsec:SUSYEqns}

The gravitational theory we consider 
is Einstein--Maxwell theory with a cosmological constant: the fields are the metric $g_{\mu\nu}$ and an Abelian gauge field $A$ with field strength $F=\rd A$, and we set the cosmological constant to $\Lambda = -3$. The resulting bulk Euclidean action is
\beq\label{eq:Action}
S = - \frac{1}{16\pi G_4}\int \left(R_g + 6 - F^2\right)\vol_g \, ,
\eeq
where $R_g$ denotes the Ricci scalar of the metric $g$, and $F^2 \equiv F_{\mu\nu}F^{\mu\nu}$. 
The equations of motion obtained from the action are
\begin{align}
\label{eq:EOMMetric}
0 & = (E_g)_{\mu\nu}\equiv (\Ric_g)_{\mu\nu} + 3g_{\mu\nu} -2\left(F_{\mu\rho}{F_{\nu}}^{\rho} - \tfrac{1}{4}F^2g_{\mu\nu}\right)  \, ,\\
\label{eq:EOMGaugeField}
0 & = \rd *_g F  \, ,
\end{align}
where (\ref{eq:EOMGaugeField}) is the Maxwell equation for the Abelian gauge field.

The action (\ref{eq:Action}) describes the bosonic sector of $\mc{N}=2$ gauged supergravity \cite{Freedman:1976aw}, and a solution to the above equations is supersymmetric if there exists a non-identically zero Dirac spinor $\epsilon$ satisfying the (generalized) Killing spinor equation coming from the vanishing of the supersymmetric gravitino variation
\beq
\label{eq:SUSY}
\left(\nabla_{\mu} - \ii A_{\mu} + \frac{1}{2}\Gamma_{\mu} + \frac{\ii}{4}F_{\nu\rho}\Gamma^{\nu\rho}\Gamma_{\mu}\right)\epsilon = 0 \, .
\eeq
Here, the Hermitian $\Gamma$ matrices generate the Clifford algebra Cliff$(4,0)$, so $\{\Gamma_\mu,\Gamma_\nu\} = 2g_{\mu\nu}$. 
Finally, we note that the supersymmetry equation \eqref{eq:SUSY} is compatible with the equations of motion \eqref{eq:EOMMetric}, \eqref{eq:EOMGaugeField} and with the Bianchi identity $\rd F=0$ for the gauge field. Specifically, the integrability condition contracted with $\Gamma^{\nu}$ leads to
\beq
\begin{split}
0 = (E_g)_{\mu\nu}\Gamma^{\nu}\epsilon + \ii \left[ \frac{1}{2}\rd F_{\mu\nu\rho}\Gamma^{\nu\rho} - (*_g\; \rd *_g F)^{\nu}\Gamma_{\mu\nu} + (*_g\; \rd F)_\mu \Gamma_* + (*_g\; \rd *_g F)_{\mu} \right]\epsilon \, ,
\end{split}
\eeq
where $E_g$ is defined in the Einstein equation \eqref{eq:EOMMetric}, and we have defined the volume element $\Gamma_*\equiv \Gamma_{1234}$.

\subsection{Reduction to a base}
\label{subsec:Reduction}

In this subsection we assume we are given a 
solution to the equations of motion \eqref{eq:EOMMetric}, \eqref{eq:EOMGaugeField} that is equipped with a 
$U(1)$ symmetry generated by a vector field $\xi$. 
This is taken to preserve both the metric and gauge field 
curvature, so $\mathcal{L}_\xi g= 0 = \mathcal{L}_\xi F$. 
The aim will be to reduce the various geometric quantities
to a three-dimensional base space of orbits of the $U(1)$ symmetry, and also obtain 
an expression for the bulk action \eqref{eq:Action}, 
evaluated on-shell. This straightforwardly generalizes the similar analysis 
in \cite{Gibbons:1979xm} for pure gravity to the 
case with an Abelian gauge field, although we will be 
careful to keep track of how various quantities 
transform under gauge transformations. 
We shall see in section \ref{sec:EvalOSAction} 
that supersymmetric solutions are always equipped 
with a canonical Killing vector $\xi$, and moreover
there is a natural gauge choice for 
the Abelian gauge field $A$. 

The assumption of a $U(1)$ symmetry acting on the 
spacetime manifold $M$ immediately 
leads to a circle fibration  
\beq\label{eq:circlebundle}
\pi:M\setminus M_0 \rightarrow B\, ,
\eeq
where 
\beq
\label{eq:M0}
M_0\equiv \{\xi=0\}\subset M
\eeq
 is the subset of $M$ where the Killing vector is zero (the fixed point set of the $U(1)$ symmetry), and $B$ is a three-dimensional base of non-trivial orbits. 
In general $B$ will be an orbifold, 
with orbifold points being images under $\pi$ of
points in $M$ with non-trivial finite isotropy subgroups 
of the $U(1)$ action.
 We may also remove small tubular neighbourhoods of radius $\varepsilon>0$ around each connected component of 
$M_0$ to obtain $M_\varepsilon\subset M$, so that $B_{\varepsilon} \equiv (M\setminus M_\varepsilon)/U(1)\subset B$ is an orbifold with boundary. 
We may then recover $B$ as the $\varepsilon\rightarrow 0$ limit.  

Next we may introduce coordinates so that the Killing vector $\xi=\partial_\psi$, where on a generic orbit $\psi$ is a local periodic coordinate with 
$\psi\sim \psi + \beta$. 
On $M\setminus M_0$ we may then write the line element for the spacetime metric $g$ as
\beq
\rd s^2 = V(\rd\psi + \phi)^2 + V^{-1} \gamma_{ij}\rd x^i \rd x^j \, .
\eeq
Here $\phi$ is a local one-form satisfying $\xi \hook \phi = 0$ and $\cL_{\xi}\phi = 0$, $V \equiv \langle \xi, \xi \rangle$ is the square norm of the Killing vector, and $\gamma$ is a metric on $B$. 
Notice that since $V$ is invariant under $\xi$, 
it descends to a strictly positive 
function 
on $B$, so that $V^{-1}$ is well-defined on $B$.
We will denote 
\beq
\label{eq:etadef}
\eta \equiv V^{-1} \xi^{\flat} = \rd\psi + \phi\, 
\eeq
where $\xi^\flat$ is the one-form dual to $\xi$ 
using the metric. The one-form $\eta$ is 
 globally defined on $M\setminus M_0$, being proportional to a global angular form for the circle bundle over $B$ in (\ref{eq:circlebundle}), while the second expression 
in (\ref{eq:etadef}) is valid only locally.  Note that under a redefinition $\psi\mapsto \psi+\lambda$, the local one-form $\phi$ transforms as $\phi \mapsto \phi - \rd\lambda$.
In addition we may define the \emph{twist} one-form by
\beq
\label{eq:twistdef}
H \equiv *_{\gamma}\rd\eta\, ,
\eeq
which is clearly conserved on $B$ by construction
\beq\label{eq:divH}
\overline{\nabla}^iH_i = *_{\gamma}\, \rd *_{\gamma}(*_{\gamma}\rd\eta) = 0 \, ,
\eeq
where $\overline{\nabla}$ is the Levi-Civita connection associated to $\gamma$ on the base.

Ultimately our aim in this section is to reduce 
the bulk action \eqref{eq:Action} to an expression 
on $B$, showing that it is naturally exact. To this end, 
we thus begin by reducing 
the Ricci tensor of the spacetime metric $g$  along the circle fibre of (\ref{eq:circlebundle}). In terms of the above 
quantities we find
\begin{align}
(\Ric_g)_{\psi \psi} &= \tfrac{1}{2}V^4 \langle H,H \rangle_{\gamma} - \tfrac{1}{2}V^2 \overline{\nabla}^2 \log V \\
(\Ric_g)_{\psi i} &= \tfrac{1}{2}\left(*_{\gamma}\rd (V^2 H) \right)_i \, , \\
\begin{split}
(\Ric_g)_{ij} &= (\Ric_\gamma)_{ij} - \tfrac{1}{2}V^{-2}\overline{\nabla}_i V \overline{\nabla}_j V + \tfrac{1}{2} \overline{\nabla}^2 \log V \, \gamma_{ij} + \tfrac{1}{2}V^2 H_iH_j \\
& \ \ \  - \tfrac{1}{2}V^2 \langle H,H \rangle_{\gamma} \gamma_{ij} \, ,
\end{split}
\end{align}
so that the scalar curvature takes the form
\beq\label{eq:Rreduced}
R_g = V R_{\gamma} + V \overline{\nabla}^2\log V - \tfrac{1}{2}V^3 \langle H, H \rangle_{\gamma} - \tfrac{1}{2} V^{-1} \langle \rd V, \rd V \rangle_{\gamma} \, .
\eeq
Here we have denoted the pointwise inner product of 
two $p$-forms $\mu$, $\nu$ on $B$ as 
$\langle \mu, \nu \rangle_{\gamma} \equiv 
 \mu_{i_1\cdots i_p}\nu^{i_1\cdots i_p}$, 
where indices are raised using the metric $\gamma_{ij}$.

Locally on $M\setminus M_0$ we may write the Abelian gauge field as
\beq
\label{eq:Agaugechoice}
A = \j \eta + a \, ,
\eeq
where we have simply decomposed this local one-form into a component along the circle fibre direction, and a transverse one-form $a$, with $\xi \hook a =0$. 
From the assumption that 
the gauge field curvature $F$ is invariant under the Killing vector, one can verify 
that it is possible to locally choose a gauge in which both $\j$ and $a$ are invariant under $\xi$, so that both descend to the base space $B$. 
In fact for supersymmetric solutions we shall see in section \ref{sec:EvalOSAction} that there is a natural gauge choice of the form 
(\ref{eq:Agaugechoice}) in which $\varphi$ is a  \emph{global} function on $B$, determined by the Killing spinor $\epsilon$, and $a$ is a local gauge one-form on $B$, so we henceforth assume this to be the case. 
This leads to the following expression for the gauge field strength
\beq\label{eq:Fdecomp}
F = \rd\j \wedge\eta + \j \, \rd\eta + f \, ,
\eeq
where $f=\rd a$. Notice that 
we may identify $\varphi=\xi \hook A$, but that
a constant gauge transformation along the $\rd\psi$ direction leads to the changes
\beq
\label{eq:GaugeTransformationField}
A \mapsto A + c\, \rd \psi \qquad \Rightarrow \qquad \j \mapsto \j + c, \qquad f\mapsto f - c\, \rd\eta \, ,
\eeq
where $c$ is a constant. We shall keep track 
of how various quantities transform under this gauge 
transformation in what follows. 
Choosing the orientation $\vol_g = V^{-1}\eta\wedge \vol_{\gamma}$, we find that
\beq\label{eq:FStarreduced}
*_gF = - V^{-1}*_{\gamma}\rd \j + \j V \eta\wedge *_{\gamma}\rd\eta + V \eta \wedge *_{\gamma}f \, ,
\eeq
which is indeed gauge invariant under \eqref{eq:GaugeTransformationField}.

\subsection{Reduction of the bulk action}
\label{subsec:ReduceAction}

Upon substitution of \eqref{eq:Rreduced} and \eqref{eq:FStarreduced} into the bulk action \eqref{eq:Action}, and further integration along the circle fibre of the $U(1)$ isometry, we find an expression for the action on the base $B$. We then add to this a Lagrange multiplier term imposing the constraint that $H$ is conserved \eqref{eq:divH}:
\beq
\label{eq:ActionB}
\begin{split}
S &= - \frac{\beta}{16\pi G_4}\int_B\bigg[ R_{\gamma} + \overline{\nabla}^2 \log V - \tfrac{1}{2}V^2 \langle H, H \rangle_{\gamma} - \tfrac{1}{2}V^{-2}\langle \rd V, \rd V \rangle_{\gamma} + 6 V^{-1}  \\
& \qquad \qquad \quad  - 2 V^{-1} \langle \rd \j, \rd\j \rangle_{\gamma}  - 2\j^2 V \langle H, H \rangle_{\gamma} - 2 \j V \langle *_{\gamma} H, f \rangle_{\gamma} - V \langle f, f \rangle_{\gamma} \\
& \qquad \qquad \quad  - \nutpot \, \overline{\nabla}^iH_i  \bigg]\vol_{\gamma} \, .
\end{split}
\eeq
This expression extends the analogue for pure Einstein gravity found in \cite{Gibbons:1979xm}. The equation of motion for the Lagrange multiplier $\nutpot$ is obviously just the constraint (\ref{eq:divH}) on $H$, whereas the equation of motion for $a$, where recall 
$f=\rd a$,  is more interesting:
\beq
\rd (\j V H + V*_{\gamma} f) = 0 \, .
\eeq
This equation implies that locally on $B$ there is an electromagnetic potential function $\empot$ such that
\beq\label{eq:EMPotential}
\rd\empot = \j V H + V*_{\gamma} f \, = \xi \hook *_gF\, ,
\eeq 
which notice is gauge invariant under \eqref{eq:GaugeTransformationField}. Here
 the second equality follows from 
 comparing with \eqref{eq:FStarreduced}. 
 In terms of the electromagnetic potential, the equation of motion for $\j$ becomes
\beq
\label{eq:EOMvarphi}
\rd *_{\gamma}\left(V^{-1}\rd\j - \empot H\right) = 0 \, ,
\eeq
which in turn tells us that we may locally write
\beq\label{eq:DefinitionVarPi}
*_{\gamma}(V^{-1}\rd\j - \empot\, H) = \rd \varpi
\eeq
for some one-form $\varpi$.

The Lagrange multiplier $\sigma$ is known 
as the \textit{nut potential} \cite{Gibbons:1979xm}. 
It also appears in the equation of motion for $H$, which 
may be expressed in terms 
 of the electromagnetic potential $\empot$ as
\beq\label{eq:NutPotential}
\rd\sigma = V^2H + 4\varphi \, \rd\nu \, .
\eeq
The origin of the name nut potential is clearer in the pure gravity case, where the last term in \eqref{eq:NutPotential} vanishes. With the introduction of an Abelian gauge field, the nut potential is not gauge invariant under \eqref{eq:GaugeTransformationField}: since $\nu, H$ and $V$ are left untouched by the gauge transformation along $\rd\psi$, we find that $\sigma$ changes as
\beq\label{eq:GaugeTransformationSigma}
\sigma \mapsto \sigma + 4c\nu \, .
\eeq

Finally, we note the equations of motion for the scalar $V$, and the Einstein equation for the metric $\gamma$ on the base:
\begin{align}
\label{eq:EOMV2}
\begin{split}
0 &= \overline{\nabla}^2 \log V - V^2 \langle H, H \rangle_{\gamma} - 6 V^{-1} + 2V^{-1} \left(\langle \rd\varphi, \rd\varphi \rangle_{\gamma} - \langle \rd\nu, \rd\nu \rangle_{\gamma} \right) \, ,
\end{split} \\
\label{eq:EOMgamma}
\begin{split}
0 &= (\Ric_{\gamma})_{ij} - \tfrac{1}{2}V^{-2}\overline{\nabla}_iV\overline{\nabla}_jV + 6 V^{-1}\gamma_{ij} + \tfrac{1}{2}V^2H_iH_j \\
& \ \ \ - 2V^{-1} \big(\overline{\nabla}_i \j \overline{\nabla}_j \j - \overline{\nabla}_i \empot \overline{\nabla}_j \empot \big)\, .
\end{split}
\end{align}
Notice that all the equations of motion are written in terms of quantities that are invariant under \eqref{eq:GaugeTransformationField}, \eqref{eq:GaugeTransformationSigma}.

Substituting these equations of motion inside the expression for the bulk Euclidean action \eqref{eq:ActionB}, we readily find an expression for the bulk \emph{on-shell} action, 
which we refer to as $I_{\rm bulk}$, 
expressed as an integral over the base:
\beq
\label{eq:OSActionB}
I_{\rm bulk} = \frac{\beta}{16\pi G_4}\int_B \rd *_{\gamma} \left( \rd\log V - \nutpot H + 4 V^{-1} \j \, \rd \j \right) \, .
\eeq
This is not immediately gauge invariant, but can be written in a more invariant way using $\varpi$
\beq
I_{\rm bulk} = \frac{\beta}{16\pi G_4}\int_B\rd \left[ 4\j\, \rd\varpi + *_{\gamma}\left( \rd\log V + \left(4\j \nu - \nutpot\right) H \right)\right] \, .
\eeq
The bracketed term contains the gauge invariant combination $4\j\empot - \nutpot$, and $\varpi$ is invariant, so the overall variation of the integrand  under \eqref{eq:GaugeTransformationField} is $\rd (4c\rd\varpi)=0$. Indeed, the variation of the integrand of \eqref{eq:OSActionB} is precisely the equation of motion for $\j$, \eqref{eq:EOMvarphi}.

The expression \eqref{eq:OSActionB} shows that 
the on-shell action is naturally exact on the base 
$B$. Importantly, we shall see in section 
\ref{sec:EvalOSAction} that for supersymmetric solutions 
there is a natural gauge in which 
both $\j$ and $\nutpot$ are \emph{global} functions 
on $B$. Since also $V>0$ is a global function, 
and $H$ a global one-form on $B$, it follows 
that we may integrate the action \eqref{eq:OSActionB} by 
parts using Stokes' theorem,  reducing to an integral over the boundary of $B$. For an asymptotically locally
Euclidean AdS spacetime, which is the case we are interested in, 
the latter consists of two types of boundary term: 
the conformal boundary at infinity, and the 
boundaries of the tubular neighbourhoods surrounding the fixed point loci of the isometry, where the  
fibration of spacetime \eqref{eq:circlebundle} degenerates. It follows that 
\eqref{eq:OSActionB} then reduces to a contribution 
from the conformal boundary, together with a 
 sum over contributions from each connected component 
 of the fixed point set of $\xi$. We shall 
 evaluate this more explicitly for supersymmetric 
 solutions in the next section.

\section{Supersymmetric solutions}
\label{sec:EvalOSAction}

\subsection{Supersymmetry equations}
\label{subsec:SUSYeqnsNEW}

In this section we shall assume we are given a 
supersymmetric
solution to the equations of motion \eqref{eq:EOMMetric}, \eqref{eq:EOMGaugeField}, meaning there 
is a non-identically zero Dirac spinor $\epsilon$ satisfying the (generalized) Killing spinor equation \eqref{eq:SUSY}. 

It will be convenient for what follows to 
introduce the charge conjugation matrix $\cC$ satisfying
\beq
\cC = \cC^* = -\cC^T \, , \qquad \cC^2 = -1 \, , \qquad \Gamma^T_\mu = \cC^{-1}\Gamma_{\mu}\cC \, .
\eeq
It was shown in \cite{Dunajski:2010uv} that 
there are no solutions to \eqref{eq:SUSY} in which 
the spinor $\epsilon$ is chiral. Thus at a generic 
point on $M$ the spinor $\epsilon$ is a 
non-chiral Dirac spinor, which in four dimensions generates an identity structure. Effectively, this corresponds to an orthonormal frame $\{\E^1,\E^2,\E^3,\E^4\}$ constructed in terms of the normalized chiral projections of the spinor
\beq
\eta_{\pm} \equiv \frac{\epsilon_{\pm}}{\sqrt{S_{\pm}}} \, , \qquad \mbox{where} \qquad  \epsilon_\pm \equiv \frac{1}{2}(1\pm \Gamma_*)\epsilon\, ,  \quad S_{\pm} \equiv \overline{\epsilon_{\pm}}\epsilon_{\pm} \, ,
\eeq
as
\beq
\label{eq:SUSYFrame}
\ii \E^3 - \E^4 \equiv \overline{\eta_-}\Gamma_{(1)}\eta_+ \, , \qquad \ii \E^1 - \E^2 \equiv \overline{\eta_-^{c}}\Gamma_{(1)}\eta_+ \, ,
\eeq
where $\epsilon^c\equiv \cC \epsilon^*$ is the charge conjugate of a spinor.
Furthermore, a Dirac spinor defines two functions $S$ and $\theta$, where $S$ is the square norm of the spinor. These
are related to the square norms $S_\pm$ of the two chiral projections via
\beq
S \equiv \overline{\epsilon}\epsilon \, , \qquad \cos^2\frac{\theta}{2} \equiv \frac{S_+}{S} \, , \qquad \sin^2\frac{\theta}{2} \equiv \frac{S_-}{S} \, .
\eeq
Notice that the frame degenerates where the spinor becomes chiral, $S_\pm=0$, that is, at $\theta=\pi, 0$, respectively, and also potentially where the spinor vanishes, $S=0$.
We may then express the standard supersymmetric bilinears in terms of the orthonormal frame and the two scalars constructed above. In particular, the following real bilinears will be relevant for us
\beq
\label{eq:SUSYBilinears}
\begin{split}
P &\equiv \overline{\epsilon}\Gamma_*\epsilon =  S \cos\theta \, , \\
K &\equiv \overline{\epsilon}\Gamma_{(1)}\epsilon = - S \sin\theta\, \E^4 \, , \\
\xi^{\flat} &\equiv -\ii\overline{\epsilon}\Gamma_{(1)}\Gamma_*\epsilon = S \sin\theta \, \E^3 \, , \\
U &\equiv \ii \overline{\epsilon}\Gamma_{(2)}\epsilon = -S (\E^{12} + \cos\theta \, \E^{34})   \, ,
\end{split}
\eeq
where note that the volume element corresponds to the orientation given by $\E^{4123}$, not $\E^{1234}$.

As is usual, from the supersymmetry equation \eqref{eq:SUSY} we can find a number of differential equations satisfied by the bilinears. In particular, we find that the vector field $\xi$, dual to $\xi^\flat$, is a Killing vector, and that
\begin{align}
\label{eq:dS}
\xi \hook *_gF &= K + \rd S \,  , \\
\label{eq:dP}
\xi \hook F &= - \rd P \, , \\
\label{eq:dK}
\rd K &= 0 \, , \\
\label{eq:dxib1}
\rd \xi^{\flat} &= -2 \left(*_gU + S*_gF + PF \right) \, .
\end{align}
Since by construction $F=\rd A$ automatically satisfies the Bianchi identity $\rd F=0$, then it follows from the equations above that $\xi$ generates a symmetry of all the bosonic fields in the solution
\beq
\label{eq:invariant}
\cL_\xi S = \cL_\xi \theta = 0 \, , \qquad \cL_\xi F = 0 \, , \qquad \cL_\xi g = 0 \, .
\eeq
Moreover, since $K$ is closed, we find that a component of Maxwell's equation (\ref{eq:EOMGaugeField}) is immediately implied
\beq
\xi \hook \rd *_gF = 0 \, .
\eeq

We conclude this subsection by mentioning a few additional points that will be relevant to our subsequent analysis. Firstly, it is not necessary for the four-dimensional spacetime $M$ 
to be a spin manifold. From (\ref{eq:SUSY}) we see that the Killing spinor $\epsilon$ has unit charge under the Abelian gauge field $A$, and so generically can be a section of 
a spin$^c$ bundle over $M$. In fact, this is precisely what happens in some of the examples we consider later: $A$ is more precisely a connection on a virtual square root line bundle $\mathcal{L}^{1/2}$, meaning 
that the periods of the globally defined curvature two-form $F=\rd A$ are in general half-integer multiples of $2\pi$, rather than the integer multiples for a standard $U(1)$ gauge field.\footnote{For further details in the current context, the interested reader is referred to Appendix D of \cite{Martelli:2012sz}.} In general we shall therefore assume that our Euclidean supergravity solutions 
are equipped with a global spin$^c$ Dirac spinor $\epsilon$, which we have shown above defines a canonical frame on the subspace of $M$ where it is non-chiral (and non-zero). 
Notice also from (\ref{eq:SUSYFrame}) that since $\ii\E^1- \E^2$ has charge two under the Abelian gauge field, it transforms as a section of the gauge line bundle $(\mathcal{L}^{1/2})^2\cong \mathcal{L}$, so more precisely this is a twisted frame. 

Next, since equation \eqref{eq:SUSY} is linear in $\epsilon$, it is possible to rescale the spinor by an arbitrary non-zero complex number, which implies a rescaling of the Killing vector $\xi$ by an arbitrary positive real number. 
Moreover, taking the gauge field $A$ to be real implies 
that the charge conjugate spinor $\epsilon^c$ satisfies the same equation \eqref{eq:SUSY}, but with $A\mapsto -A$. The latter is a symmetry of the action (\ref{eq:Action}) and equations of motion 
(\ref{eq:EOMMetric}), (\ref{eq:EOMGaugeField}).
 Constructing the Killing vector instead out of this charge
conjugated spinor, one finds that $\xi\mapsto-\xi$. 
For a  given real supersymmetric solution, there is no canonical way to choose between using $\epsilon$ or $\epsilon^c$ for its supersymmetric structure. 
From these comments, one thus expects any final result depending on the Killing vector $\xi$ to be independent of multiplying it by an 
 arbitrary real number.

Finally, we note that 
charge conjugation is also related to the choice of orientations. On the spacetime manifold $M$ we choose the orientation naturally provided by the supersymmetric frame as $\vol_g = \E^{4123}$. If instead we construct the same frame as \eqref{eq:SUSYFrame} with $\epsilon\mapsto\epsilon^c$, we find that $\E^1$ and $\E^3$ have changed sign (consistently with what was said above about the Killing vector $\xi$, given that $\E^3$ is proportional to the dual one-form $\xi^\flat$ (\ref{eq:SUSYBilinears})). Therefore, whilst the overall orientation of $M$ is the same, the individual orientations of the two tangent planes spanned by $\{\E^1, \E^2\}$ and $\{\E^3, \E^4\}$ have changed. Identifying both of these tangent planes with the complex plane $\C$, this amounts to a complex conjugation of both, while note that 
$A\mapsto -A$ also complex conjugates the Hermitian line bundle $\mathcal{L}$ on which $2A$ is a connection.

\subsection{Local form of supersymmetric solutions}
\label{subsec:SUSYSolutions}

In this subsection we proceed to further 
analyse the spinor bilinear equations presented in 
the previous subsection. This will allow 
 us to determine the local form of any real supersymmetric solution. An equivalent analysis appears in 
 \cite{Dunajski:2010uv}, but we have found it more 
 convenient to use the bilinear formalism 
developed in the present paper. In particular, 
as we shall see, certain bilinear equations play 
a key role in evaluating the on-shell action. 
 
 As already noticed, $\xi$ is a Killing vector, so 
 as in section \ref{subsec:Reduction} 
 we may introduce coordinates so that $\xi=\partial_\psi$. 
In section \ref{subsec:Reduction}  we assumed that the orbits of $\xi$ all close, so that it generates a $U(1)$ isometry of $M$.  More generally this need not be the case, but 
provided the closure of the orbits of $\xi$ in the isometry group of $M$ is a compact group, then $M$ will in fact have a torus isometry, i.e. at least $U(1)^2$. 
In this case we may approximate $\xi$ by a sequence of Killing vectors, each of which generates a $U(1)$ isometry. We shall return to this point again later. 
 From \eqref{eq:SUSYBilinears} we may write down
\beq
\E^3 = S \sin\theta (\rd\psi + \phi) \, ,
\eeq
where $\phi$, as in the previous section, is a local basic one-form. We have also seen in section \ref{subsec:SUSYeqnsNEW} that 
$\xi$ generates a full symmetry of the solution, so all functions will be independent of $\psi$. Additionally, from \eqref{eq:SUSYBilinears} and \eqref{eq:dK}, since $K$ is closed we 
may locally introduce a function $y$ satisfying
\beq
\E^4 = \frac{1}{S\sin\theta}\rd \left(\frac{1}{y}\right) \, .
\eeq
Following \cite{Dunajski:2010uv}, via an appropriate frame rotation we may choose a gauge where 
$\partial_y \hook A = 0$ . In such gauge, we may then introduce a local complex coordinate $z$ and a real function $W$ to write
\beq
\E^1 + \ii \E^2 = \frac{2\, \e^{W/2}}{y^2S \sin\theta}\rd z \, .
\eeq
From the supersymmetry equation \eqref{eq:SUSY}, we find that
\beq
\label{eq:dE1E2}
\rd (S \sin\theta (\E^1 + \ii \E^2)) - 2\ii A\wedge (S\sin\theta (\E^1 + \ii \E^2)) =  2(\E^4 - \ii \cos\theta\, \E^3 )\wedge S  (\E^1 + \ii \E^2) \, ,
\eeq
which will be useful below.

As for the gauge field, by writing it in the form \eqref{eq:Agaugechoice} and comparing its curvature with \eqref{eq:dP}, we immediately find that
\beq
\label{eq:jSUSY}
\j = S \cos\theta + c_{\j} = P +c_{\j}~,
\eeq
for some real constant $c_{\j}$. Recall here that in terms of the spinor bilinears in \eqref{eq:SUSYBilinears} we have $P\equiv \bar{\epsilon}\Gamma_*\epsilon = S\cos \theta$, so that $P$ and hence also $\varphi$ are  then manifestly global functions 
on the spacetime four-manifold $M$.  Moreover, in the gauge where $\partial_y \hook A = 0$, the components of \eqref{eq:dE1E2} fix $a$ and impose a constraint between the functions $W, S, \theta$:
\begin{align}
\label{eq:FirstConstraint}
\frac{y}{4}\partial_y W &= 1 - \frac{1}{yS \sin^2\theta} \, , \\
a_{\zbar} &= - \frac{\ii}{4}\partial_{\zbar}W \, .
\end{align}
It follows that the metric and gauge field on the spacetime respectively take the form
\begin{align}
\label{eq:LocalSUSYMetric}
\rd s^2 = S^2\sin^2\theta (\rd\psi + \phi)^2 + \frac{1}{y^4S^2\sin^2\theta}\left(\rd y^2 + 4\e^W \rd z \rd \zbar\right) \, , \\
\label{eq:LocalSUSYGaugeField}
A = \left( S \cos\theta + c_{\j}\right) (\rd\psi + \phi) + \frac{\ii}{4}\left( \partial_{z}W \, \rd z - \partial_{\zbar} W\, \rd \zbar\right) \, .
\end{align}

In order to apply the formulae derived in section \ref{subsec:Reduction} to this class of supersymmetric solutions, we should identify the terms in \eqref{eq:LocalSUSYMetric}, \eqref{eq:LocalSUSYGaugeField}. Clearly, the square norm of the Killing vector field $V$ and the metric on the base $\gamma$ are
\beq
\begin{split}
V &= S^2\sin^2\theta \, , \qquad \gamma = \frac{1}{y^4}\left( \rd y^2 + 4\e^W\rd z \rd\zbar \right) \, .
\end{split}
\eeq
Notice that the subset $M_0\equiv \{\xi=0\}$ of fixed points of the isometry is also $M_0=\{V=0\}$, which is precisely where the canonical frame degenerates, which in turn 
is where the spinor 
becomes chiral. Moreover, \eqref{eq:dS} fixes the electromagnetic potential
\beq
\label{eq:EMPotSUSY}
\empot = S - \frac{1}{y} + c_{\nu} \, .
\eeq
Note that $\nu$, just like the coordinate $y$, is \textit{a priori} only defined locally.
Recall that the gauge transformation 
\eqref{eq:GaugeTransformationField} in particular shifts $\j\mapsto\j + c$, where $c$ is a constant. Thus via an appropriate gauge transformation we may 
take $c_{\j} = 0$, obtaining
\beq
\label{eq:jSUSYgauge}
\j = S\cos\theta \, , \qquad a = \frac{\ii}{4}\left( \partial_{z}W \, \rd z - \partial_{\zbar} W \, \rd \zbar \right) \, ,
\eeq
which we refer to as the \emph{supersymmetric gauge}. 

Next, in order to determine nut potential and twist, we should find $\rd\eta$ from its definition and \eqref{eq:dxib1}:
\beq
\label{eq:SecondConstraint}
\rd\eta = V^{-2}\xi \hook (\rd\xi^{\flat}\wedge \xi^{\flat}) = 2V^{-3/2} *_{\gamma} \left[ 2\cot\theta \, \rd\left(\frac{1}{y}\right) - S\, \rd\theta\right] \, .
\eeq
In deriving this, we have compared the orientation given by supersymmetry $\vol_g = \E^{4123}$, and that chosen in the reduction to the base of the fibration $\vol_g = V^{-1}\eta \wedge \vol_{\gamma}$, to find $\vol_{\gamma} = -V^{3/2}\E^{124} = \frac{2\ii\e^W}{y^6}\rd y \wedge \rd z \wedge\rd\zbar$. The expression for $\rd\eta$ then immediately gives $H= *_{\gamma}\rd\eta$ and $\nutpot$ via \eqref{eq:NutPotential}
\beq
\label{eq:SUSYNUTpotential1}
\nutpot = 2S\j + c_{\nutpot} = 2S^2\cos\theta + c_{\nutpot} = 2SP + c_{\nutpot}\, ,
\eeq
again involving a constant $c_\sigma$ which we are going to fix in the next subsection by looking at the contribution to the on-shell action from the conformal boundary. As for $\j$, we observe that the expression for the nut potential only involves the spinor bilinears $\nutpot = 2SP + c_{\nutpot}$, so it is manifestly globally defined on $M$.

Finally, the differential equations arising from higher rank differential form spinor bilinears provide two additional constraints between the functions appearing in the metric and gauge field:
\begin{align}
\label{eq:ThirdConstraint}
\partial^2_{z\zbar}W &= - \e^W\left[\partial^2_{yy}W + \frac{1}{4}(\partial_yW)^2 + \frac{12\cos^2\theta}{y^4S^2\sin^4\theta} \right] \, , \\
\begin{split}
\label{eq:GenericDzbarzf}
\partial^2_{z\zbar}f &+ \frac{\e^W}{y^2}\bigg[f\left(f^2+2\right) - y\left(2\partial_yf+\frac{3}{2}f\partial_y W\right)\ + \\
& + y^2\left(\partial_{yy}^2f + \frac{3}{2}\partial_yW\partial_yf + \frac{3}{2}f\partial_{yy}^2W + \frac{3}{4}f(\partial_yW)^2\right) \bigg]\, = \, 0 \ .
\end{split} 
\end{align}
In the latter equation, which is introduced for completeness but not needed for the purposes of this paper, the function $f$ is defined by
\beq
f \equiv -\frac{2\cos\theta}{yS \sin^2\theta} \, .
\eeq

\subsection{On-shell action: contributions from the conformal boundary}
\label{subsec:ConformalBoundary}

In this subsection we begin our analysis of the on-shell action for supersymmetric solutions that are asymptotically locally Euclidean AdS. The main result of section 
\ref{sec:4dSUGRA} was the formula \eqref{eq:OSActionB} for the bulk contribution to the on-shell action. As usual, to obtain the full on-shell action 
 we should add to this the Gibbons--Hawking--York boundary 
term, together with the standard local boundary counterterms that holographically renormalize the action (see e.g. \cite{Emparan:1999pm, Skenderis:2002wp}). The latter are both terms at the conformal boundary 
(or more precisely near the conformal boundary, using a cut-off $\delta$ that is then removed by taking the $\delta\rightarrow 0$ limit). The bulk on-shell action 
 \eqref{eq:OSActionB} is naturally exact, and the main result of this subsection will be to show that by choosing the integration constant $c_\nutpot=0$ in the nut 
potential in \eqref{eq:SUSYNUTpotential1}, all the contributions at the conformal boundary in fact cancel. The only remaining contributions
to the on-shell action then come from the fixed points of the supersymmetric Killing vector $\xi$, which we analyse in the next subsection. 
Of course, the on-shell action is independent of the choice of $c_\nutpot$, so we may regard $c_\nutpot=0$ simply as a natural and convenient gauge choice.

Following  \cite{Farquet:2014kma, Genolini:2016ecx}, we take the conformal boundary to be at $\{y=0\}=\partial M = M_3$, where the function $1/y$ then provides a natural radial coordinate 
near to this conformal boundary. We furthermore assume that the various terms appearing in the local structure of the solution described in the previous subsection have an 
analytic expansion in $y$ in a neighbourhood of the conformal boundary at $\{ y=0\}$. Essentially the same analysis appears in  \cite{Genolini:2016ecx}, using a different set of variables, 
so here we will be brief. 
In a neighbourhood of the conformal boundary we thus write
\begin{align}
\theta(y,z,\zbar) &= \theta_{(0)}(z,\zbar) + y\, \theta_{(1)}(z,\zbar) + \frac{y^2}{2}\theta_{(2)}(z,\zbar) + \mc{O}( y^3) \, , \\
\begin{split}
W(y,z,\zbar) &= W_{(0)}(z,\zbar) + y\, W_{(1)}(z,\zbar) + \frac{y^2}{2}W_{(2)} + \mc{O}(y^3) \, , 
\end{split} \\
S(y,z,\zbar) &= \frac{1}{y}S_{(-1)}(z,\zbar) + S_{(0)}(z,\zbar) + y\, S_{(1)}(z,\zbar) + \frac{y^2}{2} S_{(2)}(z,\zbar) + \mc{O}( y^3) \, , \\
\phi (y,z,\zbar) &= \phi_{(0)}(z,\zbar) + y \phi_{(1)}(z,\zbar) + \frac{y^2}{2}\phi_{(2)}(z,\zbar) + \mc{O}( y^3 ) \, ,
\end{align}
and by imposing the constraints \eqref{eq:FirstConstraint}, \eqref{eq:ThirdConstraint}, \eqref{eq:SecondConstraint}, we find
\begin{align}
\theta &= \frac{\pi}{2} + y\, \theta_{(1)} + \frac{y^2}{2}\theta_{(2)} + \mc{O}(y^3) \, , \\
W &= W_{(0)} + y\, W_{(1)} + \frac{y^2}{2}\left( -\e^{-W_{(0)}}\partial_{z,\zbar}^2W_{(0)} - 12\theta_{(1)}^2 - \frac{1}{4}W_{(1)}^2 \right) + \mc{O}(y^3) \, , \\
\begin{split}
S &= \frac{1}{y} + \frac{1}{4}W_{(1)} + y \, \left( - \frac{1}{4}\e^{-W_{(0)}}\partial_{z,\zbar}W_{(0)} - 2\theta_{(1)}^2 \right) + \frac{y^2}{2}\bigg[ \frac{1}{8} e^{-W_{(0)}} \Big(\partial^2_{z,\zbar}W_{(0)}W_{(1)}  \\
& \ \ \ - 2 \partial^2_{z,\zbar}W_{(1)}\Big)  + \frac{1}{2} W_{(1)}\theta_{(1)}^2 - \theta_{(1)}\theta_{(2)} \bigg] + \mc{O}(y^3) \, ,
\end{split} \\
\phi &= \phi_{(0)} + y^2\, \ii ( \partial_{\zbar}\theta_{(1)}\, \rd \zbar - \partial_{z}\theta_{(1)}\, \rd z ) + \mc{O} ( y^3) \, ,
\end{align}
where $\phi_{(0)}$ is constrained to satisfy
\beq\label{eq:dphi0}
\rd\phi_{(0)} = 4\ii \e^{W_{(0)}}\theta_{(1)}\, \rd z \wedge \rd \zbar \, .
\eeq
To leading order, the metric of the four-dimensional spacetime then takes the form
\begin{align}
\rd s^2 &= \left[1 + \mc{O}(y) \right] \frac{\rd y^2}{y^2} + \frac{1}{y^2}\left[ ( \rd\psi + \phi_{(0)})^2 +4 \e^{W_{(0)}}\rd z \rd \zbar + \mc{O}(y) \right] \, ,
\end{align}
confirming that it is asymptotically locally Euclidean AdS, with defining function $y$. We may then choose a natural representative for the metric on the conformal boundary
\begin{align}
\rd s^2_3 = ( \rd\psi + \phi_{(0)})^2 +4 \e^{W_{(0)}}\rd z \rd \zbar  \, .
\end{align}
Furthermore, the bulk Abelian gauge field $A$ has leading order term $A_{(0)}$ given by
\begin{align}
A_{(0)} \equiv  \left.A\right|_{y=0} = -\theta_{(1)}(\rd \psi + a_{(0)}) + \frac{\ii}{4}\left( \partial_{z} W_{(0)}\, \rd z - \partial_{\zbar} W_{(0)}\,  \rd \zbar \right) \, .
\end{align}
Proceeding to higher orders in the expansion of the supersymmetry equations leads to additional relations between the terms in the expansion of the functions, as described in detail in \cite{Genolini:2016ecx}. The conclusion is that all the higher order terms in the series solutions are determined in terms of the boundary data, characterized by $W_{(0)}, \theta_{(1)}$ (with $\phi_{(0)}$ constrained to satisfy \eqref{eq:dphi0}), 
and the free bulk functions $W_{(1)}, \theta_{(2)}$.

The geometric structure on $M_3=\{y=0\}$ induced from the bulk supersymmetry conditions is the same as that for rigid supersymmetric backgrounds in 
 three-dimensional new minimal supergravity \cite{Closset:2012ru}. In a little more detail, $\partial_{\psi}$ restricted to the boundary coincides with the canonical 
vector field for the almost contact structure with global one-form $\rd\psi + \phi_{(0)}$, and the Abelian gauge field restricted to the boundary is identified with the non-dynamical gauge field that couples to the R-symmetry current of the field theory. Since 
we are assuming that the orbits of $\xi$ all close, the boundary manifold $M_3 = \partial M$ is a Seifert three-manifold, being the total space of an orbifold circle bundle over an orbifold Riemann surface $\Sigma$. 

Having determined the expansion of the fields near the conformal boundary, we may now evaluate the corresponding contribution to the bulk on-shell action \eqref{eq:OSActionB}, after 
integrating by parts. More precisely, we consider a cut-off spacetime $M_{\delta}$, where the radial coordinate $y$ only extends to $y=\delta>0$, rather than to the conformal boundary 
located at $y=0$. We denote the resulting boundary by $\partial M_\delta \equiv M\cap \{y=\delta\}\cong M_3$, with base $\partial B_\delta = \partial M_\delta/U(1)$. 
Applying Stokes' theorem to \eqref{eq:OSActionB}, we find that the conformal boundary contribution to the 
bulk on-shell action is
\beq\label{eq:IbulkUV}
I_{\rm bulk}^{\rm UV} = -\frac{\beta}{16\pi G_4}\int_{\partial B_{\delta}}*_{\gamma}\left( \rd \log V - \nutpot H + 4 V^{-1}\j \rd \j \right)
\eeq
The sign here is due to the direction of the normal, together with compatibility of the orientation choices associated to supersymmetry and the reduction along the circle fibre. As already pointed out, the volume element on $(B,\gamma)$ is naturally $\vol_{\gamma} = 2\ii\e^W \, \rd y \wedge \rd z \wedge \rd \zbar /y^6$. The outward-pointing unit normal to $\partial B_{\delta}$ is $n= - \rd y/y^2$, so in order to appeal to Stokes' theorem with a positive sign we should use the orientation on $\partial B_{\delta}$ given by $n^{\sharp}\hook \vol_{\gamma} = - 2\ii \e^{W}\, \rd z\wedge \rd\zbar/y^4$, but this is opposite to the natural orientation on that surface, hence the sign. We may then lift the integral in \eqref{eq:IbulkUV} back up to an integral on $\partial M_\delta$:
\beq
\begin{split}
I_{\rm bulk}^{\rm UV} &= -\frac{1}{16\pi G_4}\int_{\partial M_{\delta}}\eta\wedge *_{\gamma}\left( \rd \log V - \nutpot H + 4 V^{-1}\j \rd \j \right) \, .
\end{split}
\eeq
Next we add to the bulk action the usual Gibbons--Hawking--York term, computed using the induced metric $h$ on the cut-off hypersurface $\partial M_{\delta}$
\beq
I_{\rm GHY} = - \frac{1}{8\pi G_4}\int_{\partial M_{\delta}}K\, \vol_h \, .
\eeq
The same induced metric is used to construct the counterterm action \cite{Emparan:1999pm} cancelling the divergences present in the sum $I_{\rm bulk}^{\rm UV}+I_{\rm GHY}$:
\beq
I_{\rm ct} = \frac{1}{8\pi G_4}\int_{\partial M_{\delta}}\left( 2 + \frac{1}{2}R_h\right)\vol_h \, .
\eeq
The conformal boundary contribution $I^{\rm UV}$ to the full on-shell action is then given by the sum $I_{\rm bulk}^{\rm UV}+I_{\rm GHY} + I_{\rm ct}$, in the limit of vanishing cutoff, $\delta\rightarrow 0$. Using the expansion of the fields determined earlier, it is a straightforward computation to see that this sum reduces to
\beq\label{eq:IUV}
I^{\rm UV} = \lim_{\delta\to 0}\left( I_{\rm bulk}^{\rm UV} + I_{\rm GHY} + I_{\rm ct} \right) = c_{\nutpot}\frac{1}{8\pi G_4}\int_{\partial M}\theta_{(1)} \, \eta_{(0)}\wedge\vol_2 \, ,
\eeq
where $\vol_2 = 2\ii \e^{W_{(0)}}\, \rd z \wedge \rd\zbar$. 
Remarkably, \eqref{eq:IUV} is proportional to the constant $c_\sigma$ that appears in the nut potential in \eqref{eq:SUSYNUTpotential1}. 
On the other hand, the \emph{total} on-shell action is independent of the choice of this constant. However, clearly a natural choice is to now set 
\beq\label{eq:IUVSUSY}
c_\nutpot =0 \qquad \Rightarrow \qquad 
I^{\rm UV} = 0\, .
\eeq
With this gauge choice for the nut potential, the only contribution to the on-shell action thus comes from the fixed point set of the supersymmetric Killing vector $\xi$, and 
it is this contribution to which we now turn.

\subsection{On-shell action: contributions from the fixed points of the isometry}
\label{subsec:FixedPoints}

In this subsection we would like to evaluate the contribution to the bulk on-shell action  \eqref{eq:OSActionB} from the fixed points of the isometry generated by the 
supersymmetric Killing vector $\xi$. Recall this means that  we remove a small tubular neighbourhood $M_\varepsilon$ of radius $\varepsilon>0$ around 
(each connected component of) the fixed point set $M_0=\{\xi=0\}$. The image in the base space is then 
$B_\varepsilon=(M\setminus  M_\varepsilon)/U(1)\subset B$, which is in general an orbifold with boundary. We then wish to apply Stokes' theorem to \eqref{eq:OSActionB}, 
and evaluate the boundary contributions 
 around the fixed point set, in the limit $\varepsilon\rightarrow 0$. 

The square norm of the Killing vector $\xi$ is $V= S^2\sin^2\theta$. This vanishes where the spinor is chiral and non-zero ($\theta = 0,\pi$), and also potentially where the spinor is 
zero, where the latter is equivalent to $S=0$. It follows that the fixed points of the supersymmetric isometry precisely
correspond to the subspaces where the orthonormal frame associated to the identity structure degenerates. The connected components 
of the fixed point set of a $U(1)$ isometry must have even codimension, so in four dimensions there may only be zero-dimensional nuts, and two-dimensional bolts \cite{Gibbons:1979xm}. 
At this point we notice that in fact $S$ can never be zero anywhere, unless the isometry acts trivially on $M$: $S$ is only zero where $\epsilon$ is zero, but then 
\eqref{eq:dxib1} implies that $\rd\xi^{\flat}$ is also zero at such a point, which in turn means that the vector field is identically zero on $M$ (assuming $M$ is connected). 

It follows that on a given connected component of the fixed point set the spinor $\epsilon$ has either positive or negative chirality, corresponding to $\theta=0$, $\theta=\pi$, respectively, and 
we may thus correspondingly label isolated fixed points as nut$_{\pm}$, or connected two-dimensional fixed point sets as $\Sigma_{\pm}$, if the spinor has positive or negative chirality there, respectively.
At such points $P=\pm S$, 
and in the $\varepsilon\to 0$ limit the equation for $\rd\xi^{\flat}$ \eqref{eq:dxib1} greatly simplifies:
\beq
\label{eq:dxib2}
\rd\xi^{\flat}|_{\pm} \equiv - 2\lim_{\varepsilon\to 0}S\left( \E^{34} \pm \E^{12} \pm F + *_gF \right) \, .
\eeq
By taking the (anti-)self-dual parts at subspaces with spinors with definite chirality we find
\beq
\left( \rd\xi^{\flat}|_{\pm} \right)^{\pm} = \mp 4S F^{\pm} \, , \qquad \qquad \left(\rd\xi^{\flat}|_{\pm}\right)^{\mp}  = \mp 2\lim_{\varepsilon\to 0} S\left(\E^{12}  \pm \E^{34} \right) \, .
\eeq
Here the $\pm$ superscripts denote self-dual and anti-self-dual parts of the two-forms, respectively. The second equation above 
is particularly useful, as its square norm gives $S^2|_{\pm}$:
\beq
\label{eq:Spm}
\text{At a fixed point of chirality $\pm$:} \qquad S^2\big|_{\pm} =  \frac{1}{16}\big\langle (\rd\xi^{\flat}|_{\pm})^{\mp}, (\rd\xi^{\flat}|_{\pm})^{\mp} \big\rangle_g \, .
\eeq

\tocless\subsubsection{Contribution from a nut}

\noindent A nut is an isolated fixed point of $\xi$, and we may introduce a 
  radial geodesic distance function $\rho$ from the nut, so that the nut is at $\rho=0$. 
To leading order near the nut, the metric is simply the flat space metric
\beq
\label{eq:MetricNearNUT}
\rd s^2 = \rd\rho_1^2 + \rho_1^2 \rd\varphi_1^2 +  \rd\rho_2^2 + \rho_2^2 \rd\varphi_2^2\, ,
\eeq
where $\rho^2=\rho_1^2+\rho_2^2$. For $\rho=\varepsilon>0$ small, the induced metric on $\{\rho=\varepsilon\}$ is then approximately the round 
metric on the three-sphere of radius $\varepsilon$, $S^3_\varepsilon$.
Moreover, the supersymmetric Killing vector near the nut is then
\beq\label{eq:xidiagonal}
\xi = b_1\partial_{\varphi_1} + b_2 \partial_{\varphi_2}\, .
\eeq
Here we have identified the tangent space to the nut as $\R^4=\C\oplus\C$, where $\varphi_1, \varphi_2$ are standard polar coordinates on each copy of the complex plane $\C$ with periodicity $2\pi$, and $b_1, b_2$ are the weights of the rotations in the two two-planes. Notice that the orientations here are not unique: reversing the orientations of both two-planes is equivalent to simultaneous complex 
conjugation of both copies of $\C$, and gives the same orientation on $\R^4$, and likewise 
swapping the two two-planes also leaves the overall orientation invariant. These act as $(b_1,b_2)\mapsto (-b_1,-b_2)$ and $(b_1,b_2)\mapsto (b_2,b_1)$ on the weights, respectively.  
Without loss of generality, we may then assume that $b_1>0$, and $b_2$ may then either be positive or negative. 
Requiring that the orbits of $\xi$ close means that we can write $b_1/b_2 = p/q$, with $p,q$ coprime integers of the same signs as $b_1,b_2$, respectively, 
 so the period of a generic orbit is $\beta = 2\pi p/b_1 = 2\pi q/b_2>0$. The connected component of the boundary of the  tubular neighbourhood $\partial B_{\varepsilon}$ in the base space $B$ is then the complex weighted projective space $\Sigma_2=\mathbb{WCP}^1_{[p,|q|]}$.

Using the form of the metric \eqref{eq:MetricNearNUT} shows that the square norm $V=O(\rho^2)$ as $\rho\rightarrow 0$, and it is straightforward to see that the first term in \eqref{eq:OSActionB} does not contribute to the action in the case of a nut (see also \cite{Gibbons:1979xm}). On the other hand, the last term in the action, namely the integral of  $*_{\gamma}4\beta V^{-1}\j \, \rd\j $ over the boundary of $B$, 
can also be written using \eqref{eq:FStarreduced} as the integral of $-4\varphi \eta\wedge *F$ over the corresponding boundary in $M$, which is a three-sphere of radius $\varepsilon$. 
Now for any smooth two-form, such as $*F$, the components tangent to the three-sphere must vanish at least as fast as $O(\rho^2)$ as $\rho=\varepsilon\rightarrow 0$. 
This term hence also gives zero contribution as $\varepsilon\rightarrow 0$.
It follows that the entire contribution of a zero-dimensional fixed point to the action arises from the  term involving $\rd\eta=*_\gamma H$. On the other hand, as in \cite{Gibbons:1979xm} (or
\cite{Genolini:2016ecx}) we may easily compute
\beq
\int_{\Sigma_2} \rd\eta = \beta \int_{\mathbb{WCP}^1_{[p,|q|]}}c_1(\cL) = - \frac{\beta}{pq} \, ,
\eeq
where $\mathcal{L}$ is the orbifold line bundle over $B$ associated to the $U(1)$ fibration \eqref{eq:circlebundle}. 
The contribution  to the on-shell action of isolated nuts is hence
\beq\label{eq:OSActionnut}
I_{\rm nuts} = - \sum_{\rm nuts}\frac{\beta^2}{16\pi G_4}\sigma|_{\rm nut} \frac{1}{pq} = - \frac{\pi}{2 G_4}\sum_{\rm nuts}\sigma|_{\rm nut}\frac{1}{2b_1b_2} \, ,
\eeq
where the sign is due to the normal being $\rd\rho$. 
This shows that in order to evaluate the on-shell action contribution from a nut, we now only need an expression for the nut potential. Here supersymmetry helps, because for supersymmetric solutions the nut potential is fixed by \eqref{eq:SUSYNUTpotential1}, and the choice $c_{\nutpot}=0$ taken in the previous subsection \eqref{eq:IUVSUSY} in order to have vanishing contribution to the action from conformal infinity gives
\beq
\label{eq:SUSYNutPotential2}
\sigma = 2S\j \, .
\eeq
Since also $S\big|_{\pm} = \pm P = \pm \varphi$, using \eqref{eq:Spm}
we may hence write
\beq
I_{\rm nuts} = - \frac{\pi}{2G_4}\sum_{\rm nuts_{\pm}} \pm \big\langle (\rd\xi^{\flat}|_{\pm})^{\mp}, (\rd\xi^{\flat}|_{\pm})^{\mp} \big\rangle_g \frac{1}{16b_1b_2} \, .
\eeq
Finally, in the orthonormal frame compatible with \eqref{eq:MetricNearNUT}, \eqref{eq:xidiagonal} we have
\beq
\rd\xi^{\flat}\big|_{\rm nut} = 2\begin{pmatrix}
0 & b_1 & 0 & 0 \\
-b_1 & 0 & 0 & 0 \\
0 & 0 & 0 & b_2 \\
0 & 0 & -b_2 & 0
\end{pmatrix} \, ,
\eeq
so
\beq
\big\langle (\rd\xi^{\flat})^{\mp}, (\rd\xi^{\flat})^{\mp} \big\rangle_g\big|_{\rm nut}  = 4 (b_1\mp b_2)^2\, ,
\eeq
and we reach the final expression
\beq
\label{eq:OSActionnut2}
I_{\rm nuts} =  \frac{\pi}{2G_4}\sum_{\rm nuts_{\pm}} \mp \frac{(b_1\mp b_2)^2}{4b_1b_2} =  \frac{\pi}{2G_4}\sum_{\mathrm{nuts}_\mp} \pm \frac{(b_1\pm b_2)^2}{4b_1b_2} \, .
\eeq
Of course here the values of the weights $(b_1,b_2)$ at each nut will in general be different, although we have suppressed this in the notation.
Notice that for each nut, the expression in the sum is indeed invariant under both $(b_1,b_2)\mapsto (-b_1,-b_2)$ and $(b_1,b_2)\mapsto (b_2,b_1)$, as it must be since these are simply 
different choices of bases for the same vector field action near the nut. Moreover, it is also invariant under scaling $(b_1,b_2)\mapsto \lambda (b_1,b_2)$, which 
for $\lambda>0$ is equivalent to rescaling the spinor by a non-zero complex number. Since the overall spinor normalization is arbitrary, it again follows that 
the formula for the action had to be invariant under such a rescaling.

\

\tocless\subsubsection{Contribution from a bolt}

\noindent A bolt is a two-dimensional surface $\Sigma\subset M$. This means that the image of the boundary of a small $\xi$-invariant tubular neighbourhood around $\Sigma$ in the base $B$ is 
also a copy of $\Sigma$. We denote this as $T_\varepsilon\cong \Sigma$, which is the connected component of $\partial B_\varepsilon$ around the bolt. 

We next look at equation  \eqref{eq:dxib2}. Notice that we may always decompose a differential form $\Psi$
uniquely as $\Psi = \eta\wedge \Psi_\xi + \Psi_T$, where we identify $\Psi_\xi \equiv \xi \hook \Psi$. It then follows that the transverse part of a form $\Psi$ 
with $\mathcal{L}_\xi \Psi=0$, namely $\Psi_T=\Psi - \eta\wedge \xi\hook \Psi$, may be interpreted as a two-form on our base $B$. Moreover $\Psi=0$ if and only if both $\Psi_\xi=0$ and $\Psi_T=0$. 
For a bolt, the transverse part of $\rd \xi^{\flat}$ is zero at the bolt, since 
this is the rotation matrix for the linear action of $\xi$, which only rotates the normal directions. On the other hand, manifestly $(\E^{34})_T=0$, $(\E^{12})_T=\E^{12}$, while from 
\eqref{eq:Fdecomp} and
 \eqref{eq:FStarreduced} we have
\beq\label{eq:FT}
F_T = \varphi\, \rd \eta + f\, , \qquad (*_g F)_T = -V^{-1} *_\gamma \rd \varphi\, .
\eeq
Integrating the transverse part of  \eqref{eq:dxib2} over $T_\varepsilon$ hence gives the equation
\beq\label{eq:Tepsilon}
\lim_{\varepsilon\rightarrow 0}\int_{T_{\varepsilon}}V^{-1}*_{\gamma}\rd\j = \pm \lim_{\varepsilon\rightarrow 0}\int_{T_{\varepsilon}}(\E^{12} + F_T) \, .
\eeq
Moreover, from the spinor bilinear equations in sections \ref{subsec:SUSYeqnsNEW} and \ref{subsec:SUSYSolutions} we deduce the equation
\beq\label{eq:bill}
*_{\gamma}\rd\log V = -2S\E^{12} - 2V^{-1}\j *_{\gamma}\rd\j + 2S(\j \, \rd\eta + f) \, .
\eeq

Using these preliminary results, we may now go back to the evaluation of the on-shell action  \eqref{eq:OSActionB} for a bolt. In particular, using \eqref{eq:Tepsilon} and \eqref{eq:bill} we compute
\begin{align}\label{eq:longchain}
\frac{\beta}{16\pi G_4}\lim_{\varepsilon\to 0}\int_{B_\varepsilon} \rd *_{\gamma} (\rd \log V + 4V^{-1}\j \, \rd\j)\Big|_{\rm{bolt}} &= - \frac{\beta}{16\pi G_4}\lim_{\varepsilon\to 0}\int_{T_{\varepsilon}}\rd *_{\gamma} (\rd \log V + 4V^{-1}\j \, \rd\j)\nonumber \\
&= - \frac{\beta}{8\pi G_4}\lim_{\varepsilon\to 0}\int_{T_{\varepsilon}}\left( SF_T + S\left( \j\, \rd\eta + f \right) \right) \nonumber \\
&= - \frac{\beta}{4\pi G_4}\lim_{\varepsilon\to 0}\int_{T_{\varepsilon}} SF_T\nonumber\\
&= - \frac{\beta}{2G_4}S|_{\Sigma_{\pm}} \, \int_{\Sigma_{\pm}}c_1(F) \, .
\end{align}
Here we have used the first equation in \eqref{eq:FT} in the third line, while in the last line $c_1(F)$ is the first Chern class of the Abelian gauge bundle, and we have used that $S$ 
 is constant along the bolt. To see the latter, notice from \eqref{eq:Fdecomp} that
\beq\label{eq:dvarphi}
\rd \varphi = -F_\xi \equiv -\xi\hook F \, .
\eeq
We may restrict this equation to the bolt, and then dot with a tangent vector $\mathtt{t}$ to the bolt. 
Since by assumption $F$ is smooth everywhere, notice that $F_\xi  (\mathtt{t})$ must then tend to zero at the bolt (where $\eta$ is singular). But then
 \eqref{eq:dvarphi}  implies that $\varphi$ is constant over a bolt, and since $S\big|_{\pm} = \pm \varphi$ at a fixed point of positive/negative chirality, we deduce that also $S$ 
is constant over a bolt. Notice that an expansion in geodesic normal coordinates would give, consistently with \cite{Gibbons:1979xm}, that the $*_{\gamma}\, \rd\log V$ term in \eqref{eq:longchain} results in a contribution proportional to the area of the bolt. However, for supersymmetric solutions this area term is then cancelled by the $4V^{-1}\j \, \rd\j$ term,  leaving the topological contribution 
on the last line of \eqref{eq:longchain}.

Finally, we turn to the contribution of the middle term in \eqref{eq:OSActionB}. Again, the calculation of this is essentially the same as that in \cite{Gibbons:1979xm}, and 
is  proportional to the self-intersection number of the bolt:
\beq
\label{eq:OSActionboltY}
\left.\frac{\beta}{16\pi G_4}\lim_{\varepsilon\to 0}\int_{B_{\varepsilon}}\rd *_{\gamma}(-\sigma H)\right|_{\rm{bolt}} = \frac{\beta^2}{16\pi G_4}\sigma|_{\Sigma_{\pm}} \int_{\Sigma_{\pm}}c_1(N\Sigma_{\pm}) \, .
\eeq
Since $S$ is constant on the bolt, we may evaluate it at any point using \eqref{eq:Spm}. At a bolt, $\rd \xi^{\flat}$ is a skew-symmetric operator with rank 2, so there exists an orthonormal frame in which it has the form
\beq
\label{eq:dxibBolt}
\rd \xi^{\flat}\big|_{\rm bolt}  = 2\begin{pmatrix}
0 & 0 & 0 & 0 \\ 0 & 0 & 0 & 0 \\
0 & 0 & 0 & \kappa \\ 0 & 0 & -\kappa & 0
\end{pmatrix} \, ,
\eeq
and the ``surface gravity'' $\kappa$ is related to the period $\beta$ of a generic orbit of $\xi$  as $\kappa = 2\pi/\beta$. Thus, applying \eqref{eq:Spm} gives
\beq
\label{eq:Sbolt}
S|_{\Sigma_{\pm}} = \frac{\pi}{\beta} \, .
\eeq
Notice that $S$ is proportional to the surface gravity, consistently with the fact that it is constant on the bolt.
Finally, adding the two contributions in \eqref{eq:longchain}, \eqref{eq:OSActionboltY}, and summing over all bolts, we obtain
\beq
\label{eq:OSActionbolt1}
I_{\rm bolts} = \frac{\pi}{2G_4} \sum_{\rm bolts \ \Sigma_{\pm}}\int_{\Sigma_{\pm}}\left( -c_1(F) \pm \tfrac{1}{4}c_1(N\Sigma_{\pm}) \right) \, .
\eeq

The expression \eqref{eq:OSActionbolt1} depends on the first Chern class $c_1(F)$ of the Abelian gauge bundle, but this may in turn be related to certain topological 
invariants at a bolt $\Sigma$ by further considering the topology of $\Sigma\subset M$. Since a bolt is
a two-dimensional orientable Riemannian manifold, it can be given a complex structure and viewed as a Riemann surface $\Sigma=\Sigma_g$ of genus $g$. 
Notice this involves a choice of orientation.  Complex line bundles over a compact Riemann surface are in one-to-one correspondence with $H^2(\Sigma_g;\Z) \cong \Z$, and the (group) isomorphism is given by the first Chern number of the bundle. Thus, we may use unambiguously the notation $\mc{O}(n)$ for the line bundle over $\Sigma_g$ with degree $n$. In particular, 
having fixed a choice of orientation for the bolt, in order to agree with the given orientation on $M$ this also fixes an orientation for the normal bundle $N\Sigma$ of 
$\Sigma$ in $M$. We may then write $N\Sigma=\mc{O}(Y)$, where the integer $Y=\int_\Sigma c_1(N\Sigma)\in \Z$ is the self-intersection number of $\Sigma$ in $M$. Similarly, 
the tangent bundle of the bolt is $T\Sigma=\mc{O}(2-2g)$, where $\Sigma=\Sigma_g$ has genus $g$.

From the above discussion it follows that topologically $TM|_{\Sigma}\cong \cO(2-2g)\oplus \cO(Y)$ may be written as a direct sum of two complex line bundles, and from this we may then compute the 
 chiral spin bundles of $M$ restricted to the bolt, which we denote as $\mc{S}_\pm$. One finds
\beq
\begin{split}\label{eq:Spmbolt}
\mc{S}_+ &= \mc{O}\left(  - (1-g) + \frac{Y}{2} \right) \oplus \mc{O}\left(  (1-g) - \frac{Y}{2} \right) \, , \\
\mc{S}_- &= \mc{O}\left( - (1-g) - \frac{Y}{2} \right) \oplus \mc{O}\left(  (1-g) + \frac{Y}{2} \right)  \, .
\end{split}
\eeq
Of course, as spin bundles here we should also keep track of the inequivalent spin structures, classified by $H^1(M;\Z_2)$. This amounts 
to the choice of a $\Z_2$ principal bundle, or equivalently a flat complex line bundle $\mathscr{S}$ with $\Z_2$-valued holonomy, that arise since both $\pm 1\in \mathrm{Spin}(4)$ map to
the identity in $SO(4)$.  More precisely, we should then tensor $\mc{S}_\pm$ in \eqref{eq:Spmbolt} with $\mathscr{S}$, although as we shall see below these different choices 
of spin structure play no role in the following argument. Independently of this, the bundles in \eqref{eq:Spmbolt}
 in general only exist as genuine vector bundles when $M$ is a spin manifold. However, as briefly mentioned
in section \ref{subsec:SUSYeqnsNEW}, generally $M$ is spin$^c$. This means that $2A$ is a connection on a genuine complex line 
bundle $\cL$ over $M$, with $c_1(\cL)\ \mbox{mod 2}=w_2(M)\in H^2(M;\Z_2)$ reducing mod 2 to the second Stiefel--Whitney class of $M$.  
Indeed, we see from \eqref{eq:SUSY} that the connection acting on the spinor is  twisted by $A$. 
This ensures that a corresponding twisting by $\cL^{1/2}$ leads to well-defined spin$^c$ spinor bundles:
\beq
\label{eq:BundlesNearBolt}
 \begin{split}
\mc{S}_+\otimes \cL^{1/2} &= \mc{O}\left(  - (1-g) + \frac{m+Y}{2} \right) \oplus \mc{O}\left(  (1-g) + \frac{m-Y}{2} \right) \, , \\
\mc{S}_-\otimes \cL^{1/2} &= \mc{O}\left( - (1-g) + \frac{m-Y}{2} \right) \oplus \mc{O}\left(  (1-g) + \frac{m+Y}{2} \right)  \, .
\end{split}
\eeq
where we have denoted $\mathcal{L}|_\Sigma=\mc{O}(m)$. These correspond to the $2+2=4$ components of a spin$^c$ Dirac spinor.

As we have shown, the spinor is necessarily chiral on (a connected component of) the fixed point set, but with non-zero constant norm. This means that there is a nowhere-zero section of one of the two pairs of line bundles in \eqref{eq:BundlesNearBolt}, which means that one of the complex line bundles is just the trivial bundle at the bolt. In order to find out which one it is, we can refer to the form of $U$ in terms of the supersymmetric vierbein in \eqref{eq:SUSYBilinears}, which near a bolt$_{\pm}$, reduces to
\beq
U|_{\Sigma_{\pm}} = - \lim_{\varepsilon\to 0}S (\E^{12} \pm \E^{34}) \, ,
\eeq
which we can verify to be consistent with the following projection conditions\footnote{Recall here that the supersymmetric vierbein is not that used to define the $\Gamma$ matrices, as remarked below \eqref{eq:SUSYBilinears}.} on the spinor
\beq
\ii \Gamma_{12}\epsilon = \epsilon \, , \quad \ii \Gamma_{34}\epsilon = \mp \epsilon \, .
\eeq
With our choice of conventions, the only complex line bundles in \eqref{eq:BundlesNearBolt} that could admit a non-vanishing section satisfying the projection conditions near the bolt are the second 
summands in each line, so in the limit $\varepsilon\to 0$, near a bolt$_{\pm}$, the non-zero spinor component is a section of
\beq
\mc{O}\left( (1-g) + \frac{m\mp Y }{2} \right) \overset{!}{=} \mc{O}(0) \, .
\eeq
This in turn then constrains the gauge bundle near a bolt in terms of the local topology 
\beq
\cL|_{\Sigma_{\pm}} = \mc{O}(m) = \mc{O}(\pm Y) \oplus \mc{O}(-2(1-g)) = N\Sigma_{\pm}^{\pm 1} \oplus T\Sigma_{\pm}^{-1} \, ,
\eeq
or, equivalently,
\beq
\label{eq:FluxAndGeometry}
\int_{\Sigma_{\pm}}c_1(F) = \int_{\Sigma_{\pm}}\frac{\pm c_1(N\Sigma_{\pm}) - c_1(T\Sigma_{\pm})}{2} \, .
\eeq

With this result, we can go back to substitute the first Chern class of the gauge bundle through the bolt in \eqref{eq:OSActionbolt1}, whence
\beq
\label{eq:OSActionbolt2}
I_{\rm bolts_{\pm}} = \frac{\pi}{2G_4}\sum_{\rm bolts \ \Sigma_{\pm}}\int_{\Sigma_{\pm}} \left(\tfrac{1}{2}c_1(T\Sigma_{\pm}) \mp \tfrac{1}{4}c_1(N\Sigma_{\pm}) \right) \, .
\eeq
Summing \eqref{eq:OSActionnut2} and \eqref{eq:OSActionbolt2} leads to the expression for the on-shell action already quoted in the Introduction, equation \eqref{Iintro}.

\

\section{Examples}
\label{sec:Examples}

As already remarked in the Introduction, there are a number of explicitly known supersymmetric solutions to $\cN=2$ gauged supergravity. In this section, we illustrate some features of the general formulae given above by considering concrete examples.

\subsection{\texorpdfstring{AdS$_4$}{AdS4}\label{sec:AdS}}

The simplest example we consider is AdS$_4$, realised as a hyperbolic ball foliated by three-spheres
\beq
\label{eq:AdS4}
\rd s^2 = \frac{\rd r^2}{r^2+1} + r^2 \left( \rd\vartheta^2 + \cos^2\vartheta \, \rd\varphi_1^2 + \sin^2\vartheta \, \rd\varphi_2^2 \right) \, ,
\eeq
with a trivial Abelian gauge field, $A=0$.
The radial coordinate is $r\in[0,\infty)$, while the $S^3$ is viewed as a torus fibration over the interval, with $\vartheta\in[0,\tfrac{\pi}{2}]$ parametrizing the interval, and $\varphi_1, \varphi_2 \in [0,2\pi)$ 
parametrizing the torus.\footnote{The change of variable $r^2 = 4y^2/(1-y^2)^2$ takes \eqref{eq:AdS4} to the standard metric on the Poincaré ball
\beq
\rd s^2 = \frac{4}{(1-y^2)^2}\left( \rd y^2 + y^2 \rd\Omega_3^2 \right) \, .
\eeq}
This is a maximally supersymmetric solution to the equations of motion of our theory and can be given a number of supersymmetric structures. Indeed solving the (generalized) Killing spinor equation on this background leads to a spinor that depends on four complex numbers. By appropriately choosing them, we may construct supersymmetric Killing vectors as in \eqref{eq:SUSYBilinears} that have the form
\beq
\xi^{\pm} = \partial_{\varphi_1}\pm \partial_{\varphi_2} \, .
\eeq
Such vectors generate an  isometric torus action on AdS$_4$ and have a nut at $r=0$ (in the nomenclature of \cite{Gibbons:1979xm}, respectively a nut or an anti-nut). The corresponding spinors, in the vierbein where $\e^1 = r \cos\vartheta \, \rd\varphi_1$, $\e^2 = r \sin\vartheta \, \rd\varphi_2$, $\e^3 = r\, \rd\vartheta$, $\e^4 = \rd r/\sqrt{r^2+1}$ are
\beq
\epsilon^+ = \frac{1}{\sqrt{2}}\begin{pmatrix}
-\sinh\frac{\arcsinh r}{2} \, \e^{\ii (-\vartheta  + \varphi_1 + \varphi_2)/2} \\
\sinh\frac{\arcsinh r}{2} \, \e^{\ii (\vartheta + \varphi_1 + \varphi_2)/2} \\
-\ii\cosh\frac{\arcsinh r}{2} \, \e^{\ii (-\vartheta  + \varphi_1 + \varphi_2)/2} \\
\ii\cosh\frac{\arcsinh r}{2} \, \e^{\ii (\vartheta  + \varphi_1 + \varphi_2)/2}
\end{pmatrix} \, , \
\epsilon^- = \frac{1}{\sqrt{2}}\begin{pmatrix}
\cosh\frac{\arcsinh r}{2} \, \e^{\ii (\vartheta  - \varphi_1 + \varphi_2)/2} \\
-\cosh\frac{\arcsinh r}{2} \, \e^{\ii (-\vartheta - \varphi_1 + \varphi_2)/2} \\
\ii\sinh\frac{\arcsinh r}{2} \, \e^{\ii (\vartheta  - \varphi_1 + \varphi_2)/2} \\
-\ii\sinh\frac{\arcsinh r}{2} \, \e^{\ii (-\vartheta  - \varphi_1 + \varphi_2)/2}
\end{pmatrix} \, .
\eeq
Here we have chosen the following form of the $\Gamma$ matrices generating Cliff$(4,0)$
\beq
\label{eq:GammaMatrices}
\Gamma_i = \begin{pmatrix}
0 & \sigma_i \\ \sigma_i & 0
\end{pmatrix} \, , \qquad \Gamma_4 = \begin{pmatrix}
0 & \ii \identity_2 \\
- \ii \identity_2 & 0
\end{pmatrix} \, \qquad \Gamma_* \equiv \Gamma_{1234} = \begin{pmatrix}
\identity_2 & 0 \\ 0 & - \identity_2
\end{pmatrix} \, ,
\eeq
where $\sigma_i$ are the Pauli matrices, and the charge conjugation matrix is
\beq
\cC = \begin{pmatrix}
\ii \sigma_2 & 0 \\ 0 & - \ii \sigma_2
\end{pmatrix} \, .
\eeq
It is immediately clear from the above form of the spinors that near the nut the spinors $\epsilon^{\pm}$ have negative/positive chirality.\footnote{One should bear in mind that a rotation of the $S^3$ frame is needed to make it regular at the origin, so that the spinors are then manifestly regular at the origin in such a frame.} Therefore, corresponding to the relative signs of the two circle actions in the torus we have either a nut$_{-}$ or a nut$_{+}$ (the nut is a nut$_{-}$ and the anti-nut is a nut$_{+}$). It is then immediate to see that applying the formula \eqref{Iintro} leads in both cases to the action of AdS$_4$
\beq
I_{{\rm AdS}_4} = \frac{\pi}{2G_4} \, .
\eeq

However, AdS$_4$ admits a number of different supersymmetric structures with corresponding Killing vectors constructed out of the generalized Killing spinors. For instance, a different supersymmetric structure was found in \cite{Martelli:2011fu} and reviewed in \cite{Farquet:2014kma}: this corresponds to a non-trivial Abelian instanton gauge field and a Killing spinor leading to the generic circle action
\beq
\xi = b_1 \partial_{\varphi_1} + b_2 \partial_{\varphi_2} \, ,
\eeq
which subsumes the above case, corresponding to $b_1/b_2 = \pm 1$.
As shown in \cite{Farquet:2014kma}, the resulting solution is in fact regular only if $b_1/b_2>0$ or if $b_1/b_2 = -1$, and the chirality analysis can be shown to carry through: if $b_1/b_2>0$, we have a nut$_-$. Thus, application of \eqref{Iintro} leads to the same result as \cite{Farquet:2014kma}.

\subsection{Self-dual solutions}
\label{subsec:SD}

As we have just reviewed, starting with the standard hyperbolic metric \eqref{eq:AdS4} on AdS$_4$, one can turn on an instanton gauge field that modifies the standard Killing spinors 
on AdS$_4$, 
and hence also the supersymmetric Killing vector. 
In fact requiring the gauge field to be anti-self-dual is natural in four dimensions, and leads to interesting classes of solutions related to integrable systems, described in general in \cite{Dunajski:2010zp}.
Under the assumption that the topology of the spacetime is that of a four-ball, with a nut at the origin (as for the explicit solutions in section \ref{sec:AdS} above), the value of the on-shell action has been computed in terms of the weights of a torus action at the nut in \cite{Farquet:2014kma}, and further analysis has been performed in \cite{Genolini:2016ecx}, so here we will be succinct.

Requiring the gauge field to be anti-self-dual implies that
\beq
\rd (\j - \empot) = 0 \quad \Rightarrow \quad \j = \empot + k \, ,
\eeq
where $k\in\R$ is a constant, which we choose to be zero (assuming that spacetime is path-connected). For supersymmetric solutions, for which the electromagnetic potential $\empot$ and $\j$ are fixed by \eqref{eq:EMPotSUSY} and \eqref{eq:jSUSYgauge}, this also means that
\beq
\cos\theta = 1 - \frac{1}{Sy} \, .
\eeq
Therefore, the square norm of the Killing vector is $V=(2Sy-1)/y^2$. In order to more easily compare with \cite{Farquet:2014kma}, we introduce the function
\beq
\cV \equiv \frac{1}{2Sy-1} \, .
\eeq
Using this notation, it is immediate to show that the metric has the form
\beq
\label{eq:MetricASD}
\rd s^2 = \frac{1}{y^2}\left[ \cV^{-1}\eta^2 + \cV (\rd y^2+ 4\e^W\, \rd z \rd\zbar) \right] \, ,
\eeq
whereas \eqref{eq:FirstConstraint} and \eqref{eq:SecondConstraint} reduce to
\beq
\begin{split}
\cV &= 1 - \frac{1}{2}y\, \partial_y W \, , \\
\rd\phi &= -\ii \rd y \wedge \left( \partial_z \cV\, \rd z - \partial_{\zbar}\cV \, \rd\zbar\right) - 2\ii \, \partial_y \left( \cV \e^W \right) \, \rd z \wedge \rd\zbar \, .
\end{split}
\eeq
and \eqref{eq:ThirdConstraint} becomes the $SU(\infty)$ Toda equation
\beq\label{eq:Toda}
\partial_{z\zbar}^2W + \partial^2_y\e^W = 0 \, .
\eeq
This matches precisely the structure described in \cite{Dunajski:2010uv, Farquet:2014kma}.\footnote{There is one caveat: the supersymmetric Killing vector field there has the opposite sign to ours, so $\eta^{\rm here} = - \eta^{\rm there}$. This is consistent with both papers considering anti-self-dual gauge fields.} Thus, the entire solution is described by the solution to a single partial differential equation.
More geometrically, requiring the gauge field to be anti-self-dual forces the $U(1)$ stress-energy tensor in \eqref{eq:EOMMetric} to vanish (one uses the Schouten identity), so the metric is Einstein, and the Weyl tensor of \eqref{eq:MetricASD} is anti-self-dual as well. For reference, four-dimensional self-dual Einstein  manifolds are considered the four-dimensional analogues of higher-dimensional quaternionic K\"ahler manifolds, and hence are sometimes referred to as such.

In terms of the metric function $\cV$, we may rewrite
\beq
\label{eq:ASDcostheta}
\cos\theta = \frac{1-\cV}{1+\cV} \, ,
\eeq
which we use to determine whether at the fixed point of the isometry the spinor has positive or negative chirality. As in \cite{Farquet:2014kma, Genolini:2016ecx}, we restrict to supersymmetric solutions where the spacetime has the topology of $\R^4$ with a nut at the origin. As proved in the original papers, given a generic toric Killing vector field
\beq
\xi = b_1\partial_{\varphi_1} + b_2 \partial_{\varphi_2} \, ,
\eeq
the metric is regular everywhere outside the nut provided $b_1/b_2>0$ (in which case the nut is at a finite radial coordinate $y=y_{\rm NUT} = 1/(b_1+b_2)$), or $b_1=-b_2$ (in which case the radial coordinate diverges at the location of the nut). In the first case, if $\rho^2$ denotes the geodesic distance from the nut, near the nut we have $\cV\sim \rho^{-2}$ (as required in order to have a smooth metric), so equation \eqref{eq:ASDcostheta} tells us that the fixed point is a nut$_-$. The value of the on-shell action given by our formula \eqref{Iintro} matches precisely that computed in \cite[eq. (1.2)]{Farquet:2014kma}
\beq\label{eq:ballaction}
I = \frac{\pi}{2G_4}\frac{(b_1+b_2)^2}{4b_1b_2} \, .
\eeq
If, instead, $b_1=-b_2$, in order to have a smooth metric near the nut we need $\cV\sim \rho^2$, so \eqref{eq:ASDcostheta} gives that the spinor at the fixed point has positive chirality. The on-shell action in this case is given by
\beq
I = \frac{\pi}{2G_4} \, ,
\eeq
which again matches the result of \cite{Farquet:2014kma}.

\subsection{Non-self-dual \texorpdfstring{$1/4$}{1/4} BPS}
\label{subsec:14BPS}

Non-self-dual solutions that are $1/4$ BPS have been constructed in \cite{Martelli:2012sz} and generalized in \cite{Toldo:2017qsh}. The spacetime has the topology of a complex line bundle over a Riemann surface of genus $g$, so that we may write $M=\cO(-p)\rightarrow \Sigma_g$, where $p\in\Z_{>0}$. The metric and gauge field are given by
\beq
\label{eq:14BPSMetricandGauge}
\begin{split}
\rd s^2 &= \frac{r^2-s^2}{\Omega(r)}\rd r^2 + (r^2-s^2)(\sigma_1^2 + \sigma_2^2) + 4s^2\frac{\Omega(r)}{r^2-s^2}\sigma_3^2 \\
A &= -\frac{(4s^2-\kappa)(r^2+s^2)+4Qrs}{2(r^2-s^2)} \sigma_3 \, ,
\end{split}
\eeq
where $\kappa=+1,0,-1$ denotes the curvature of $\Sigma_g$ and the one-forms $\sigma_i$ are
\begin{itemize}[leftmargin=0.4cm]
\item for $\kappa=+1$, $g=0$ and $\Sigma_0 = S^2$
\begin{gather}
\sigma_1 = \cos\tau\, \rd \vartheta + \sin\tau\sin\vartheta\, \rd \phi \, , \quad \sigma_2 = - \sin\tau \, \rd\vartheta + \cos\tau\sin\vartheta\, \rd\phi \, , \quad \sigma_3 = \rd\tau + \cos\vartheta\, \rd\phi \, , \nn \\
\sigma_1^2 + \sigma_2^2 = \rd\vartheta^2 + \sin^2\vartheta\, \rd\phi^2 \, ,
\end{gather}
\item for $\kappa=0$, $g=1$ and $\Sigma_1 = T^2$
\begin{gather}
\sigma_1 = \sin\tau \, \rd \vartheta + \cos\tau\, \rd\phi \, , \quad \sigma_2 = -\cos\tau\, \rd\vartheta + \sin\tau \, \rd\phi \, , \quad \sigma_3 = \rd\tau - \vartheta \, \rd \phi \, , \nn \\
\sigma_1^2 + \sigma_2^2 = \rd\vartheta^2 + \rd\phi^2 \, ,
\end{gather}
\item for $\kappa=-1$, $g>1$ and $\Sigma_g$ a Riemann surface obtained by compactifying $\mathbb{H}^2$
\begin{gather}
\sigma_1 = \sin\tau\, \rd \vartheta + \cos\tau\sinh\vartheta\, \rd \phi \, , \quad \sigma_2 = - \cos\tau \, \rd\vartheta + \sin\tau\sinh\vartheta\, \rd\phi \, , \quad \sigma_3 = \rd\tau - \cosh\vartheta\, \rd\phi \, , \nn \\
\sigma_1^2 + \sigma_2^2 = \rd\vartheta^2 + \sinh^2\vartheta \, \rd\phi^2 \, .
\end{gather}
\end{itemize}
The function $\Omega(r)$ has the form
\beq
\Omega(r) = (r^2-s^2)^2 + (\kappa - 4s^2)(r^2-s^2) - 4sQ\, r + \frac{1}{4}(4s^2-\kappa)^2 - Q^2 \, ,
\eeq
and is required to be positive.

 In order to write down the expression for the Killing spinor, we introduce a frame
\beq
\label{eq:14BPSFrame}
\begin{split}
\e^1 &= \sqrt{r^2-s^2}\, \sigma_1 \, , \quad \e^2 = \sqrt{r^2-s^2}\, \sigma_2 \, , \\
\e^3 &= 2s\sqrt{\frac{\Omega(r)}{r^2-s^2}}\, \sigma_3 \, , \quad \e^4 = \sqrt{\frac{r^2-s^2}{\Omega(r)}}\, \rd r \, .
\end{split}
\eeq
The Killing spinor satisfying \eqref{eq:SUSY} with our choice of $\Gamma$ matrices \eqref{eq:GammaMatrices} and in our frame is
\beq
\label{eq:14BPSspinor}
\epsilon = \begin{pmatrix}
0 \\ \sqrt{\frac{(r-r_3)(r-r_4)}{r-s}}\, \chi_{(0)} \\ 0 \\ \ii \sqrt{\frac{(r-r_1)(r-r_2)}{r+s}}\, \chi_{(0)}
\end{pmatrix} \, ,
\eeq
where $r_{1,2,3,4}$ are the four roots of $\Omega$ and $\chi_{(0)}$ is a complex constant which we set to $\sqrt{s}$ to simplify expressions (recall that we may always renormalize the Killing spinor by a complex constant without affecting the result). Since the frame \eqref{eq:14BPSFrame} introduced above is only locally defined, so is the expression \eqref{eq:14BPSspinor} for the spinor: a more detailed discussion of the global regularity of the solution can be found in the appendices to \cite{Martelli:2012sz, Toldo:2017qsh}.

The supersymmetric Killing vector obtained as a bilinear from \eqref{eq:14BPSspinor} is $\xi = \partial_{\tau}$, and it is clear by looking at the form of the metric \eqref{eq:14BPSMetricandGauge} that the norm of the supersymmetric Killing vector vanishes at the roots of $\Omega$. We have a conformal boundary as $r\to\infty$ and the metric is complete if it collapses smoothly at the largest root of $\Omega$. If $Q$ is such that $\Omega$ has a double root, that is $Q=\pm \tfrac{1}{2}(4s^2-\kappa)$, then necessarily this is at $r=s$, we are only allowed to consider the $g=0$ case, and we reduce to the self-dual Taub--NUT--AdS solution, which is covered by the discussion in the 
 previous subsection \cite{CEJM}. However, generically the solution is not self-dual, and considering a neighbourhood of the simple largest root $r_0$ shows that the fixed point locus of $\xi$ is a two-dimensional bolt $\Sigma_g$. In order to have a smooth collapse of the plane perpendicular to the bolt, we need to impose
\beq
\label{eq:Regularity14BPS}
\frac{r_0^2-s^2}{|\Omega'(r_0)|}= \frac{2s}{\mb{p}} \, ,
\eeq
where
\beq
\mb{p} = \frac{p}{|g-1|} \ \ \ \text{for $g\neq 1$} \, , \qquad \mb{p} = p \ \ \ \text{for $g=1$} \, .
\eeq
This shows concretely that the spacetime has the topology of an $\cO(-p)$ fibration over the bolt $\Sigma_g$. In particular, the coordinate $\tau$ parametrizes the polar angle 
direction in the complex line fibre, having period $4\pi |g-1|/p$ when $g\neq 1$, and period $4\pi/p$ when $g=1$ \cite{Martelli:2012sz, Toldo:2017qsh}.

Imposing the regularity condition \eqref{eq:Regularity14BPS}, together with some algebra, implies that there are two branches of solutions with different largest root of $\Omega$ and different value for $Q$; we refer to them as positive and negative branches and label the bolts as $\Sigma_g^{\pm}$ (note that here $\pm$ does not refer to the chirality of the spinor at the bolt, although see the discussion further below). Moreover, for different values of the integer $p$, each branch will only exist for certain ranges of the deformation parameter $s$. The moduli space of solutions is quite intricate and we would not do justice to it here, so we refer the interested reader to \cite{Martelli:2012sz} and \cite{Toldo:2017qsh} for a more extensive discussion. 

In the following, we move to the local analysis of the $U(1)$ fibration by the supersymmetric Killing vector orbits. Knowing the Killing vector and its norm allows us to find the metric on the base
\beq
\gamma = 4 s^2 \left( \rd r^2 + \Omega(r)(\sigma_1^2 + \sigma_2^2) \right) \, ,
\eeq
and the twist vector
\beq
H = - \frac{1}{2s\Omega(r)}\rd r \, .
\eeq
Comparing the expression \eqref{eq:14BPSMetricandGauge} with \eqref{eq:Agaugechoice} gives
\beq
\j = -\frac{(4s^2-\kappa)(r^2+s^2)+4Qrs}{2(r^2-s^2)} + c_{\j} \, ,
\eeq
and because of the gauge choice in $c_\j$, \eqref{eq:GaugeTransformationField} imposes that $f = - c_\j\, \rd\eta$. Because of the form of the metric, $\empot$ and $\nutpot$ may be simply found by integrating along the radial coordinate. From \eqref{eq:EMPotential} we find
\beq
\empot = - s \frac{(4s^2-\kappa)r+2sQ}{r^2-s^2} + c_{\nu} \, ,
\eeq
and from \eqref{eq:NutPotential}, we find
\beq
\begin{split}
\nutpot &= \frac{2s}{(r^2-s^2)^2} \bigg[4 s^2 \left(Q^2 r + 2 Q s \left(r^2+s^2\right) - r \left(r^2-3 s^2\right) \left(r^2+s^2\right)\right)\\
& \ \ \ -4 \kappa  r s \left(Q r+s^3+r^2 s\right)+\kappa^2 r^3 \bigg] + 4c_{\j}\nu + c_{\nutpot} \, ,
\end{split}
\eeq
where we notice the natural appearance of the $4c_\j \empot$ term, consistently with \eqref{eq:GaugeTransformationSigma}.

At this point, we should fix the gauge choices $c_\j, c_{\nutpot}$: to do so, we additionally require that the functions satisfy the relations \eqref{eq:jSUSYgauge} and \eqref{eq:SUSYNutPotential2},  characteristic of supersymmetric solutions in the supersymmetric gauge. This fixes
\beq
\label{eq:14BPSGaugeChoice}
c_\j = - \frac{\kappa}{2} \, , \qquad c_{\nutpot} = 2(\kappa c_{\empot} - 4Qs^2) \, .
\eeq
By performing standard holographic renormalization, we readily see that the contribution from conformal infinity to the on-shell action vanishes for the choice \eqref{eq:14BPSGaugeChoice}, consistently with the fact that in the supersymmetric gauge the contribution from conformal infinity is zero \eqref{eq:IUVSUSY}.

Now we can finally check the validity of our computation near the fixed point locus: near the bolt in the positive/negative branch, the spinor \eqref{eq:14BPSspinor} has negative/positive chirality, so the bolt $\Sigma_g^{\pm}$ is a bolt$_{\mp}$. Computing the flux of the gauge field through the bolt, this confirms the relation between gauge bundle and geometry \eqref{eq:FluxAndGeometry}, since
\beq
\int_{\Sigma_g^{\pm}} c_1(F) = \frac{\pm p - 2(1-g)}{2} = \int_{\Sigma_g^{\pm}}\frac{\mp c_1(N\Sigma^{\pm}_g) - c_1(T\Sigma^{\pm}_g)}{2} \, ,
\eeq
where $T\Sigma_g \cong \cO(2-2g)$ and $N\Sigma_g \cong \cO(-p)$ (recall that in the notation of \eqref{eq:FluxAndGeometry}, $\Sigma^{\pm}_g$ is a bolt $\Sigma_{\mp}$!). Moreover, this concretely shows that the gauge field is generically a connection on a virtual unitary line bundle, making the spinor a spin$^c$ spinor -- only for even $p$ is $A$ a connection on an honest $U(1)$ bundle.

It is then easy to find that the only non-vanishing skew-eigenvalue of $\rd\xi^{\flat}$ is the surface gravity of the bolt $\kappa = - \frac{\mb{p} }{2}$ (in the normalization of \eqref{eq:dxibBolt}), and that consistently with \eqref{eq:Sbolt}, at leading order we have $S = |\kappa|/2$. 
We can then apply \eqref{Iintro} to obtain the value of the on-shell action
\beq
\label{eq:14BPSAction}
I_{\pm} = \frac{\pi}{2G_4}\left(\mp \frac{p}{4} + (1-g) \right) \, ,
\eeq
consistently with the results obtained previously in the literature \cite[eqs (3.73) and (3.74)]{Toldo:2017qsh}. Notice that the analysis of the supersymmetric structure performed here clarifies further the appearance of the branches of solutions, which appeared in the original literature out of considering the smoothness of the metric, and justifies their different on-shell action.

\subsection{Non-self-dual \texorpdfstring{$1/2$}{1/2} BPS}
\label{subsec:12BPS}

In addition to the solutions considered in the previous section that preserve $1/4$ of the supersymmetries, solutions preserving $1/2$ of the supersymmetries have been found in \cite{Martelli:2012sz}. The ansatz has $SU(2)\times U(1)$ isometry, and the local form of the metric and gauge field is very similar to \eqref{eq:14BPSMetricandGauge}
\beq
\label{eq:12BPSMetricandGauge}
\begin{split}
\rd s^2 &= \frac{r^2-s^2}{\Omega(r)}\rd r^2 + (r^2-s^2)(\sigma_1^2 + \sigma_2^2) + 4s^2\frac{\Omega(r)}{r^2-s^2}\sigma_3^2 \\
A &= -\frac{s\sqrt{4s^2-1} (r^2+s^2)+2Qrs}{(r^2-s^2)} \sigma_3 \, ,
\end{split}
\eeq
but now $\sigma_i$ are the $SU(2)$ left-invariant one-forms
\beq
\sigma_1 + \ii \sigma_2 = \e^{-\ii\tau}\left(\rd\vartheta + \ii \sin\vartheta\, \rd\phi \right) \, , \qquad \sigma_3 = \rd\tau + \cos\vartheta\, \rd\phi \, ,
\eeq
and the function $\Omega$ reads
\beq
\Omega(r) = (r^2-s^2)^2 + (1 - 4s^2)(r^2-s^2) - 2Q\sqrt{4s^2-1}\, r + s^2(4s^2-1) - Q^2 \, .
\eeq
Because the solution has the same local form (at least formally, $\Omega$ being different), we choose the same vierbein as \eqref{eq:14BPSFrame}. In that frame, the spinor satisfying \eqref{eq:SUSY} has the form
\beq
\epsilon = \begin{pmatrix}
\sqrt{\frac{(r-r_3)(r-r_4)}{r-s}} \chi^{(+)} \\
\sqrt{\frac{(r-r_1)(r-r_2)}{r-s}} \chi^{(-)} \\
\ii \sqrt{\frac{(r-r_1)(r-r_2)}{r+s}} \chi^{(+)} \\
\ii \sqrt{\frac{(r-r_3)(r-r_4)}{r+s}} \chi^{(-)}
\end{pmatrix} \, ,
\eeq
where $r_i$ are the roots of $\Omega$, and
\beq
\begin{pmatrix}
\chi^{(+)} \\
\chi^{(-)}
\end{pmatrix} = 
\begin{pmatrix}
\cos\tfrac{\vartheta}{2} \, \e^{\ii(\tau + \phi)/2} & - \sin\tfrac{\vartheta}{2} \, \e^{\ii(\tau-\phi)/2} \\
\gamma  \sin\tfrac{\vartheta}{2} \, \e^{-\ii(\tau-\phi)/2} & \gamma \cos\tfrac{\vartheta}{2} \, \e^{-\ii(\tau+\phi)/2}
\end{pmatrix}
\begin{pmatrix}
\mathtt{p} \\ \mathtt{q}
\end{pmatrix} \, ,
\eeq
where $\gamma \equiv \ii(2s+\sqrt{4s^2-1})$ and $(\mathtt{p}, \mathtt{q})\in \C^2\setminus\{0\}$. The supersymmetric Killing vector field constructed with this spinor according to \eqref{eq:SUSYBilinears} is
\beq
\begin{split}
& \xi \equiv \partial_{\psi} = -2\Big\{(2s + \sqrt{4s^2-1}) \Big[ 2\Im \left(\mathtt{p}\overline{\mathtt{q}}\e^{\ii\phi} \right) \partial_{\vartheta} + \left(|\mathtt{p}|^2 - |\mathtt{q}|^2 + 2\Re\left(\mathtt{p}\overline{\mathtt{q}}\e^{\ii\phi}\right)\, \cos\vartheta \right) \partial_{\phi}\Big] \\
& \ + \left[\left( \frac{1}{2s}-2s-\sqrt{4s^2-1} \right) \left( |\mathtt{p}|^2 + |\mathtt{q}|^2\right) - 2\Re\left( \mathtt{p}\overline{\mathtt{q}}\e^{\ii\phi} \right) \left(2s+\sqrt{4s^2-1}\right)\csc\vartheta \right]\partial_{\tau} \Big\} \, .
\end{split}
\eeq
Apart from an irrelevant overall normalization factor, this agrees with \cite[eq. (5.23)]{Farquet:2014kma}. Notice that the supersymmetric Killing vector does not simply correspond to the Killing vector $\partial_{\tau}$ generating the Hopf fibration. Since the form of the metric is the same as in the previous subsection, the solution has spherical bolts for $\partial_\tau$ at the largest root of $\Omega$, as was the case in the $1/4$ BPS solutions. These bolts correspond to the base spaces of the fibration that determines the topology of spacetime: for the self-dual case, the largest root of $\Omega$ is $r=s$ and the solution is just Taub--NUT--AdS, whereas in general the topology is that of a bundle $\cO(-p)\rightarrow S^2$. In this sense, these solutions are referred to as ``bolt'' solutions in the literature. However,
 we are interested in the fixed points of the isometry generated by the supersymmetric Killing vector, so in the nomenclature of the current paper they should be referred to as ``nut'' solutions. 
Indeed, $\xi$ vanishes at the two poles of the $S^2$ when one of the two parameters $\mathtt{p}, \mathtt{q}$ is chosen to be zero. In other words, for $(\mathtt{p},\mathtt{q})=(1,0)$ or $(0,1)$, the supersymmetric Killing vector has  nuts at the two poles of the $S^2$ bolt of the Killing vector $\partial_{\tau}$.

Regularity of the metric near the largest root of $\Omega$ again imposes a constraint on $Q$ analogous to \eqref{eq:Regularity14BPS}. As in the previous case, there are two branches of solutions, labelled positive and negative, and an intricate moduli space of their existence depending on the value of the deformation parameter $s$ and the self-intersection number $p$. Moreover, the moduli spaces of the $1/4$ BPS and $1/2$ BPS supersymmetric solutions intersect non-trivially, making the problem of matching the value of the on-shell action particularly interesting. As before, we refer the interested readers to the original paper \cite{Martelli:2012sz} for more details. 

Examining the Killing spinor near the poles, we see that for the choice $(\mathtt{p},\mathtt{q})=(1,0)$ we have a nut of $\pm$ type near the north/south pole for the positive/negative branch of solutions, and for the choice $(\mathtt{p},\mathtt{q})=(0,1)$ we have a nut of $\mp$ type near the north/south pole for the positive/negative branch of solutions, but the value of the on-shell action does not depend on the choice of spinor (within a branch).

In the $(\mathtt{p},\mathtt{q})=(1,0)$ case, the weights of the Killing vector near the north pole are
\beq
b_1 = - 4s - 2\sqrt{4s^2-1} \, , \qquad b_2 = \frac{p}{2s}
\eeq
for both positive and negative branch of solutions, and near the south pole
\beq
b_1 = 4s + 2\sqrt{4s^2-1} \, , \qquad b_2 = \frac{p}{2}\left( \frac{1}{s}-8s-4\sqrt{4s^2-1} \right) \, ,
\eeq
again for both branches. Applying \eqref{Iintro} then leads to
\beq
\label{eq:12BPSAction}
I_{\pm} = \frac{\pi}{2G_4}\left[1 \pm \frac{2\sqrt{4s^2-1}}{sp}\left(s^2 - \frac{p^2}{16} \right)\right] \, ,
\eeq
which matches \cite[eq. (4.29)]{Martelli:2012sz}. Similarly, in the $(\mathtt{p},\mathtt{q})=(0,1)$ case, the weights of the Killing vector near the north pole are
\beq
b_1 = 4s + 2\sqrt{4s^2-1} \, , \qquad b_2 = \frac{p}{2}\left( \frac{1}{s}-8s-4\sqrt{4s^2-1} \right)
\eeq
and near the south pole
\beq
b_1 = - 4s - 2\sqrt{4s^2-1} \, , \qquad b_2 = \frac{p}{2s} \, .
\eeq
Again, applying \eqref{Iintro} leads to \eqref{eq:12BPSAction} (the weights are exchanged between the two poles, but so is the chirality of the spinor near the nuts).

\subsection{General \texorpdfstring{$\cO(-p)\rightarrow S^2$}{O(-p)->S2}}
\label{subsec:ToricOp}

The two solutions considered in the previous two subsections share the same topology $M=\mc{O}(-p)\rightarrow S^2$, where $p\in \Z_{>0}$ and with both admitting a $U(1)^2$ torus action 
which contains the isometry generated by the supersymmetric Killing vector field $\xi$.  
On the other hand, our main formula \eqref{Iintro} for the action of a solution only requires knowledge of the action of $\xi$ on $M$, together with the chirality data (determining 
certain signs) at the fixed points of $\xi$. In this section we compute the action for \emph{any} supersymmetric solution on $M=\mc{O}(-p)\rightarrow S^2$, 
\emph{assuming the solution exists}, and show that the 1/4 BPS and 1/2 BPS results in sections~\ref{subsec:14BPS} and \ref{subsec:12BPS} arise as special cases. 
As well as recovering known results very simply, we are then also able to give the actions of solutions that have not (yet) been found in closed form, 
but of course we do need to assume the solutions actually exist. We discuss this further in section \ref{sec:Conclusions}.

Four-manifolds with a $U(1)^2$ torus action have been classified in  \cite{OrlikRaymond}. In fact the latter results were used more recently, in a related context, 
in \cite{Calderbank:2002gy}, where a brief review may be found. In the case of $M=\mc{O}(-p)\rightarrow S^2$ the obvious $U(1)^2$ action 
is moreover compatible with a symplectic (and indeed K\"ahler) structure. Using the notation of sections \ref{subsec:14BPS} and \ref{subsec:12BPS},
 we let 
$\partial_\phi$ be the lift of the vector field that rotates the $S^2$ zero section, fixing the north and south poles at $\vartheta=0$, $\pi$, 
 and $\frac{2}{p}\partial_\tau$ be the vector field that rotates the complex line fibre with weight one.\footnote{Note 
that correspondingly $\frac{p}{2}\tau$ has period $2\pi$.} We then introduce the following 
basis for the $U(1)^2$ action:
\beq\label{eq:changetoricbasis}
\partial_{\psi_1} = \partial_\phi + (1-\frac{p}{2})\frac{2}{p}\partial_\tau\, , \qquad \partial_{\psi_2} = \frac{2}{p}\partial_\tau\, .
\eeq
The vector fields $\partial_{\psi_1}$, $\partial_{\psi_2}$ generate an effective action of the torus on $M=\mc{O}(-p)\rightarrow S^2$, for any (non-zero) $p$.\footnote{Of course 
any $SL(2,\Z)$ transformation of this basis will also suffice. However, notice that the shift in the expression for $\partial_{\psi_1}$ by the half-integer $p/2$ when $p$ is odd 
is required because in this case a single orbit of $\partial_\phi$ does not in fact  close on $M$. In physical language, this is because for odd $p$ the complex line bundle $\mc{O}(-p)$ 
is a half-integer spin representation, with $p=1$ in particular being the chiral spin bundle of $S^2$. In this case, a single rotation of the base 
only induces a half rotation of the fibre.}
Moreover, this is an isometric action for a natural K\"ahler structure, where we view $M=\C^3//U(1)_{1,1,-p}$ as a K\"ahler quotient of 
$\C^3$ by $U(1)$ with weights $(1,1,-p)$ on $(z_1,z_2,z_3)\in \C^3$. Stated more physically, $M$ arises as the vacuum moduli space of the gauged linear
sigma model (GLSM) with three complex scalar fields with $U(1)$ charges $(1,1,-p)$. There is then a corresponding moment 
map $\mu:M\rightarrow \R^2$, and the image of this is given by the shaded region in the figure below.

\

\

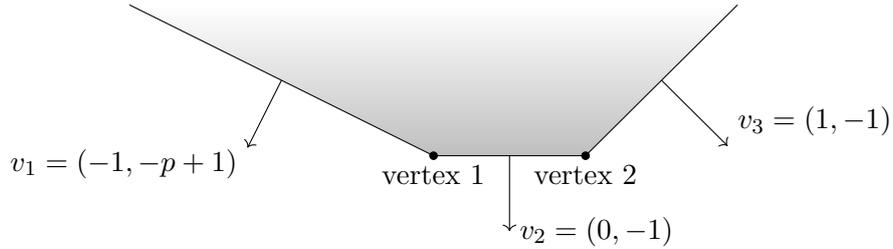
\medskip
\begin{center}
\begin{tikzpicture}
	\shade[bottom color=gray!50, top color=white] (-5,2) -- (-1,0) -- (1,0) -- (3,2);
	\draw (-5,2) -- (-1,0) node[draw, circle, inner sep=1pt, fill]{} -- (1,0) node[draw, circle, inner sep=1pt, fill]{} -- (3,2);
	\draw[->] (-3,1) -- (-3.447,0.106);
	\draw[->] (0,0) -- (0,-1);
	\draw[->] (2,1) -- (2.87,0.13);
	\node[left] at (-3.447,-0.106) {$v_1 = (-1,-p+1)$};
	\node[right] at (0,-1) {$v_2 = (0,-1)$};
	\node[above right] at (2.87,0.13) {$v_3 = (1,-1)$};
	\node[below] at (-1,0) {$\mbox{vertex 1}$};
	\node[below] at (1,0) {$\mbox{vertex 2}$};
\end{tikzpicture}
\end{center}
\medskip

Geometrically, $\mu(M)=P\subset \R^2$ is a non-compact, convex polytope. The preimage $\mu^{-1}(p)$ for $p$ in the interior $P_{\mathrm{int}}$ of $P$ is a copy of $T^2=U(1)^2$, 
and indeed $\mu^{-1}(P_{\mathrm{int}})\cong T^2 \times P_{\mathrm{int}}$ is a dense open subset of $M$. However, along the boundary $\partial P$ different  
$U(1)$ subgroups degenerate. Specifically, the pre-image under $\mu$ of each edge of the polytope is a fixed point set of the $U(1)\subset U(1)^2$ specified by the 
normal vector $v_a\in \Z^2$ to the edge -- this is a key property of the moment map image for symplectic toric manifolds. The finite edge 
with normal vector $v_2=(0,-1)$ is precisely the image under $\mu$ of the $S^2$ zero section of $M=\mc{O}(-p)\rightarrow S^2$, with the vertices 
at each end corresponding to the north and south poles. These vertices are then precisely the images of points of $M$ which are fixed under the entire $U(1)^2$ action. 

This toric diagram allows us to immediately  write down the weights of the torus action at the two vertices:
\beq
\begin{split}
\mbox{Weights at vertex 1}:  &  \qquad u_1^{(1)} = (-p+1,1) \, , \ \ \   u_2^{(1)} = (1,0) \, , \\
\mbox{Weights at vertex 2}:  & \qquad u_1^{(2)} = (-1,0) \, , \ \ \  \ \  \quad  u_2^{(2)} = (1,1) \, .
\end{split}
\eeq
Geometrically, these weights are simply the primitive outward pointing edge vectors of $P$ at each vertex, respectively. 
With this notation in hand, we may now write down a general toric Killing vector as
\beq\label{eq:xiMp}
\xi = a_1 \partial_{\psi_1}+ a_2\partial_{\psi_2} = (a_1,a_2)~,
\eeq
where $a_1,a_2\in \R$ (not both zero) determine the choice of supersymmetric Killing vector field on $M$. 
The weights of $\xi$ on each factor of $\C\oplus \C=TM_{p}$ at a fixed point/vertex $p$ are then simply 
$\xi \cdot u_1^{(i)}$, $\xi \cdot u_2^{(i)}$, respectively, where here $i=1,2$ labels the two different fixed points/vertices. 
Thus we may write down
\beq
\label{eq:ToricWeightsnuts}
\begin{split}
\mbox{vertex 1}:  &\qquad  \left( b_1^{(1)}, b_2^{(1)} \right) =\left( (-p+1)a_1 + a_2, a_1 \right) \, , \\
\mbox{vertex 2}:  & \qquad \left( b_1^{(2)}, b_2^{(2)} \right) =\left( -a_1, a_1 + a_2 \right) \, .
\end{split}
\eeq

Assuming that such a supersymmetric solution exists on $M=\mc{O}(-p)\rightarrow S^2$, with supersymmetric Killing vector 
given by \eqref{eq:xiMp}, the action of such a solution given by \eqref{Iintro} also depends on the 
chiralities associated to the two vertices. We denote these as $\kappa_1,\kappa_2 \in\{\pm 1\}$, respectively. 
Then for generic $a_1,a_2\in \R$ the fixed points of $\xi$ are precisely the two vertices, labelled by $i=1,2$, and so from 
\eqref{Iintro} we may write down the action
\beq
\begin{split}
I  & = I_p(\kappa_1,\kappa_2;a_1,a_2) =  \left[\sum_{i=1}^2 -\kappa_i\frac{\left(b_1^{(i)}-\kappa_i b_2^{(i)}\right)^2}{4b_1^{(i)}b_2^{(i)}}\right]\frac{\pi}{2G_4}\label{eq:IMpgeneral}\\
& =  \left[\frac{\mathcal{Q}_p(\kappa_1,\kappa_2;a_1,a_2)}{4 a_1(a_1+a_2) (a_1 (p-1)-a_2)}\right]\frac{\pi}{2G_4}\, ,
\end{split}
\eeq
where we have defined the homogeneous cubic polynomial
\beq
\begin{split}
&\mathcal{Q}_p(\kappa_1,\kappa_2;a_1,a_2)  =\bigg\{a_1^3 \left[\kappa_1 \left(p^2-2 p+2\right)+2 (\kappa_2+2) (p-1)\right]+a_2^3 (\kappa_1-\kappa_2) \\ & + a_1^2 a_2 (p-2) [\kappa_1 (p-2)  +2 (\kappa_2+2)]+
 a_1 a_2^2 [\kappa_1 (3-2 p)+\kappa_2 (p-3)-4]\bigg\}\, .
\end{split}
\eeq
The action $I_p(\kappa_1,\kappa_2;a_1,a_2)$ in \eqref{eq:IMpgeneral} depends on the choice of 4-manifold $M=\cO(-p)\rightarrow S^2$, via the integer $p\in \Z_{\geq 0}$, the assignment of chiralities 
$\kappa_i\in\{\pm 1\}$ to each of the two vertices (fixed points of the torus action), and also on the choice of 
supersymmetric Killing vector \eqref{eq:xiMp}, via the coefficients $a_1,a_2\in \R$. 

From this general result, we may immediately recover both the 1/4 BPS and 1/2 BPS results in the previous two subsections. 
Notice first that the 1/4 BPS solution is a degenerate case of the above analysis, where the entire $S^2$ zero section 
of $M=\cO(-p)\rightarrow S^2$ is fixed by $\xi$. This immediately requires us to take $\kappa_1=\kappa_2$, since these are 
precisely the chiralities at the two poles of the $S^2$ zero section, which must be the same since the spinor is now chiral 
over the entire $S^2$. Setting $\kappa_1=\kappa_2=\kappa$, the action simplifies to
\beq\label{eq:IMp14general}
I_p(\kappa,\kappa;a_1,a_2) = \left[\frac{a_1^2 \left(\kappa p^2+4 p-4\right)+a_1 a_2 (p-2) (\kappa p+4)-a_2^2 (\kappa p+4)}{4 (a_1+a_2) (a_1 (p-1)-a_2)}\right]\frac{\pi}{2G_4}\, .
\eeq
On the other hand, the 1/4 BPS Killing vector precisely rotates the complex line fibre of $M=\cO(-p)\rightarrow S^2$, and from \eqref{eq:changetoricbasis}, \eqref{eq:xiMp} 
we see this means 
setting $a_1=0$, so that $\xi\propto \partial_\tau$. This gives
\beq
I = I_p(\kappa,\kappa;0,a_2\neq 0) = \left( \frac{\kappa p}{4}+1\right) \frac{\pi}{2G_4}\, .
\eeq
This correctly reproduces the 1/4 BPS bolt action $I_\pm$ in \eqref{eq:14BPSAction}, where recall we should set $\kappa=\mp 1$ and the genus $g=0$ for $S^2$.
Notice we have recovered this formula as a limit of the general nut fixed point in the first line of \eqref{eq:IMpgeneral}, in the limit where the Killing vector 
$\xi$ develops an $S^2$ bolt, rather than two nuts. Conversely, we may regard \eqref{eq:IMp14general} as the 1/4 BPS action with \emph{generic} 
choice of toric Killing vector, specified by $a_1/a_2$ and the choice of chirality $\kappa$, while the explicitly known solution has $a_1=0$. 
To date such a solution is not known explicitly, or even known to exist, but assuming it \emph{does} exist, its action is given by \eqref{eq:IMp14general}. 
Notice that such a solution will necessarily have only $U(1)^2$ as isometry, rather than the $SU(2)\times U(1)$ isometry of the 1/4 BPS solution with $a_1=0$. 

Next we may recover the 1/2 BPS solution result in section \ref{subsec:12BPS}. Recall this had a nut$_-$ and a nut$_+$, so we now set $\kappa_1=-\kappa_2=\kappa$. 
Using the general form of the Killing vector $\xi$ in that section, with the choice $(\mathtt{p},\mathtt{q})=(1,0)$, it is straightforward to read off the weights 
$a_1,a_2$ in \eqref{eq:xiMp}, where recall the basis is defined by \eqref{eq:changetoricbasis}. One finds
\beq
a_1= -2(2s+\sqrt{4s^2-1})\, , \qquad a_2 = -2\left(\frac{p}{4s}-2s-\sqrt{4s^2-1}\right)\, .
\eeq
From \eqref{eq:IMpgeneral} we then compute
\beq
\begin{split}
I  & = I_p\left(\kappa,-\kappa;-2(2s+\sqrt{4s^2-1}), -2\left(\frac{p}{4s}-2s-\sqrt{4s^2-1}\right)\right) \\
& = \left[1 - \frac{2\kappa\sqrt{4s^2-1}}{sp}\left(s^2 - \frac{p^2}{16} \right)\right] \frac{\pi}{2G_4}\, .
\end{split}
\eeq
This precisely agrees with the 1/2 BPS action $I_\pm$ in \eqref{eq:12BPSAction} on setting $\kappa=\mp 1$ for the two branches, respectively. 

\subsection{More general topologies}
\label{subsec:top}

It is straightforward to generalize the computation in the above subsection to \emph{any} four-manifold $M$ with a $T^2$ action. Specifically, one can use the description 
of such four-manifolds in \cite{OrlikRaymond}, and reviewed in \cite{Calderbank:2002gy}. Provided $M$ is simply-connected, which notice we have assumed already in our 
supergravity analysis, the quotient space $M/T^2$ is topologically a polygon. As in the previous subsection, the edges of this polygon are labelled by a coprime pair of integers 
$v_a=(m_a,n_a)\in \Z^2$,  specifying the  circle subgroup $U(1)\subset T^2$ that fixes the corresponding $T^2$-invariant two-manifold in $M$. In particular, vertices of the 
polygon corresponds to points of $M$ that are fixed under the $T^2$ action, with finite edges between a pair of vertices corresponding to $T^2$-invariant two-spheres. 
However, in this general setting there is not necessarily any convexity property of the polygon, while as mentioned in the previous subsection when there is a compatible 
symplectic structure $P=M/T^2$ is naturally a convex polytope, with normal vectors to the edges of $P$ given precisely by the $v_a$. Rather than attempt a general analysis, in this subsection 
we present some further simple examples, and also make some general comments on properties of the action formula \eqref{Iintro}.

For the $M=\mc{O}(-p)\rightarrow S^2$ examples in the previous subsection, the conformal boundary is a Lens space $M_3=\partial M = L(p,1)$. There are various ways 
to see this, but one method again uses some standard toric geometry. Consider the linear map that sends
\beq\label{eq:linmap}
(1,0) \mapsto v_1=(-1,-p+1)\, , \qquad (0,1)\mapsto v_3=(1,-1)\, .
\eeq
Here $v_1$ and $v_3$ are normals to the non-compact edges of the polytope $P$, which geometrically correspond to the complex line fibres $\C$ over the north and  poles of the 
$S^2$ zero section. The linear map \eqref{eq:linmap}  sends $\Z^2\rightarrow \Z^2$, and the kernel of the induced map of tori $T^2=\R^2/T^2\rightarrow \R^2/T^2$ 
is generated by $(\tfrac{1}{p},\tfrac{1}{p})$, since $(\tfrac{1}{p},\tfrac{1}{p})\mapsto (0,-1)\in \Z^2$. Blowing down the zero section, with normal
 vector $v_2$, then gives the singular space $\C^2/\Z_p$, where the $\Z_p$ action is $\C^2 \ni (z_1,z_2)\mapsto (\omega_p z_1,\omega_p z_2)$, with
$\omega_p=\mathrm{e}^{2\pi \ii /p}$ a primitive $p$th root of unity. 

We can instead consider fillings of different Lens spaces 
\beq
L(p,q)= S^3/\Z_p\, , \qquad (z_1,z_2)\mapsto (\omega_p z_1,\omega_p^q z_2)\, ,
\eeq
where $p$ and $q$ are coprime integers with $p>q>0$, and we identify $S^3$ as the unit sphere in $\C^2$ with coordinates $(z_1,z_2)$.  
The $L(p,q)$ are toric three-manifolds, in the sense that the $\Z_p$ quotient commutes with the standard $T^2$ action on $\C^2\supset S^3$. 
The minimal resolution of the corresponding complex singularity $\C^2/\Z_p$ is well-known, and the toric data of the polygon $P$ referred to above is closely related 
to a continued fraction expansion of $q/p$ -- see, for example, \cite{Calderbank:2002gy}. Here we present the simplest example, namely $L(3,2)$. 

\

\

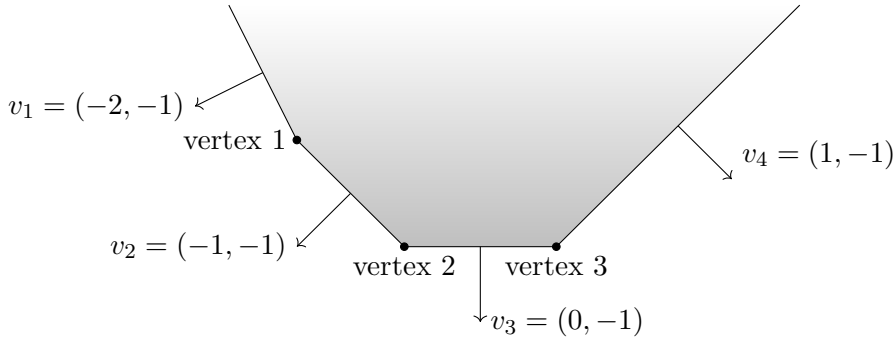
\medskip
\begin{center}
\begin{tikzpicture}
	\shade[bottom color=gray!50, top color=white] (-3.309,3.203) -- (-2.414,1.414) -- (-1,0) -- (1,0) -- (4.203,3.203);
	\draw (-3.309,3.203) -- (-2.414,1.414) node[draw, circle, inner sep=1pt, fill]{} -- (-1,0) node[draw, circle, inner sep=1pt, fill]{} -- (1,0) node[draw, circle, inner sep=1pt, fill]{} -- (4.203,3.203);
	\draw[->] (-2.861,2.308) -- (-3.755,1.861);
	\draw[->] (-1.707,0.707) -- (-2.414,0);
	\draw[->] (0,0) -- (0,-1);
	\draw[->] (2.602,1.602) -- (3.309,0.895);
	\node[left] at (-3.755,1.861) {$v_1 = (-2,-1)$};
	\node[left] at (-2.414,0) {$v_2 = (-1,-1)$};
	\node[right] at (0,-1) {$v_3 = (0,-1)$};
	\node[above right] at (3.309,0.895) {$v_4 = (1,-1)$};
	\node[left] at (-2.414,1.414) {$\mbox{vertex 1}$};
	\node[below] at (-1,0) {$\mbox{vertex 2}$};
	\node[below] at (1,0) {$\mbox{vertex 3}$};
\end{tikzpicture}
\end{center}
\medskip

In this case the vectors $v_a$ for the (K\"ahler) resolution are given by $v_1=(-2,-1)$, $v_2=(-1,-1)$, $v_3=(0,-1)$, $v_4=(1,-1)$. 
The kernel of the map of tori $T^2=\R^2/T^2\rightarrow \R^2/T^2$ generated by 
\beq
(1,0) \mapsto v_1=(-2,-1)\, , \qquad (0,1)\mapsto v_4=(1,-1)\, ,
\eeq
is this time generated by $(\tfrac{1}{3},\tfrac{2}{3})$, identifying the boundary three-manifold as $L(3,2)$. 
Notice that although $L(3,1)$ and $L(3,2)$ are homeomorphic, via the map which complex conjugates the 
second factor in $\C\oplus \C = \C^2$ (so $z_2\mapsto \bar{z}_2$), this also changes the complex structure 
of the transversely holomorphic foliation generated by the supersymmetric Killing vector $\xi$ on $M_3=\partial M$.  
There are three vertices of the polytope, corresponding to fixed points of the $T^2$ action on $M$, with weights (outward pointing edge vectors)
\beq
\begin{split}
\mbox{Weights at vertex 1}:  &  \qquad u_1^{(1)} = (-1,2) \, , \ \ \   u_2^{(1)} = (1,-1) \, , \\
\mbox{Weights at vertex 2}:  & \qquad u_1^{(2)} = (-1,1) \, , \ \  \    u_2^{(2)} = (1,0) \, ,\\
\mbox{Weights at vertex 3}:  & \qquad u_1^{(3)} = (-1,0) \, , \ \ \   \,  u_2^{(3)} = (1,1) \, .
\end{split}
\eeq
There are two finite edges of the polytope $P$, with normal vectors $v_2$ and $v_3$, which correspond to two $T^2$-invariant two-spheres which intersect 
at a point (corresponding to vertex 2).
Again writing the supersymmetric Killing vector as
\beq\label{eq:xiagain}
\xi = a_1 \partial_{\psi_1}+ a_2\partial_{\psi_2} = (a_1,a_2)~,
\eeq
it is straightforward to compute the action of a supersymmetric solution on this four-manifold, provided it exists of course. 
We take the chiralities of the vertices to be all equal, so $\kappa_1=\kappa_2=\kappa_3=\kappa\in\{\pm 1\}$. Then for 
generic $a_1,a_2\in \R$ we may use \eqref{Iintro} to write down the action
\beq\label{eq:I32action}
\begin{split}
I_{L(3,2)}(\kappa,\kappa,\kappa;a_1,a_2) & = \left[\sum_{i=1}^3 -\kappa \frac{(\xi \cdot u_1^{(i)}-\kappa \xi \cdot u_2^{(i)})^2}{4 \xi \cdot u_1^{(i)}\, \xi \cdot u_2^{(i)}}\right]\frac{\pi}{2G_4}\\
&= \left[\frac{2 a_1^2 (1 +3\kappa)-2 a_1 a_2 (1 +3\kappa)-a_2^2 (4  +9\kappa)}{4 (a_1-2 a_2) (a_1+a_1)}\right] \frac{\pi}{2G_4}\, .
\end{split}
\eeq
This can be compared with the minimal filling of $L(3,1)$ in the previous subsection, where the topology is $M=\mc{O}(-3)\rightarrow S^2$ and there is a single 
blown up two-sphere. The action with chiralities of the $T^2$ fixed points both equal to $\kappa$ is given by setting $p=3$ in \eqref{eq:IMp14general}, namely 
\beq
I_{L(3,1)}(\kappa,\kappa;a_1,a_2) = \left[\frac{a_1^2(8+9\kappa)+a_1a_2(4+3\kappa) - a_2^2(4+3\kappa)}{4(2a_1-a_2)(a_1+a_2)}\right] \frac{\pi}{2G_4}\, .
\eeq
Of course, it would be remarkable to reproduce these formulae from a dual localization calculation in SCFT -- we discuss this further in section \ref{sec:Conclusions}.

Another interesting question to address is the general behaviour of the action \eqref{Iintro} under blowing up. Recall that topologically this means 
replacing a neighbourhood of the origin in $\R^4$ by $\mc{O}(-1)\rightarrow S^2$, where both have boundary given by $S^3$. Moreover, this blowing 
up is compatible with the obvious $T^2$ action, which fixes the origin of $\R^4\cong \C\oplus \C$. In terms of the ``toric'' description of 
$M$ in \cite{OrlikRaymond}, we may then blow up the vertices of the polygon, where the blow-up simply replaces the vertex by a finite 
edge. Thus, let $p$ be a vertex, with neighbouring edges that meet at that vertex having labels $v=(m,n)$, $v'=(m',n')\in \Z^2$. 
Then the blow-up introduces a new edge with label
\beq
v_0 = (m+m',n+n')\, .
\eeq

\

\

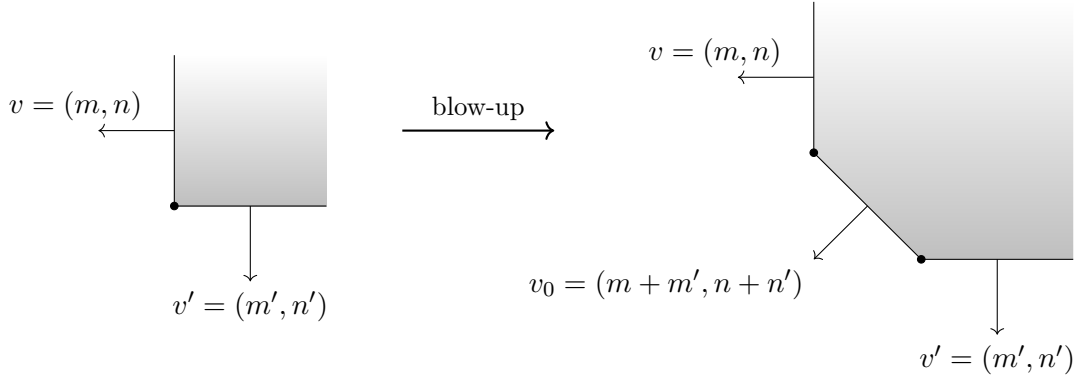
\medskip
\begin{center}
\begin{tikzpicture}
	\shade[bottom color=gray!50, top color=white] (-4,2) -- (-4,0) -- (-2,0) -- (-2,2);
	\draw (-4,2) -- (-4,0) node[draw, circle, inner sep=1pt, fill]{} -- (-2,0) -- (-2,0);
	\draw[->] (-4,1) -- (-5,1);
	\draw[->] (-3,0) -- (-3,-1);
	\node[above] at (-5.3,1) {$v = (m,n)$};
	\node[below] at (-3,-1) {$v' = (m',n')$};
	\draw[->,thick] (-1,1) -- (1,1);
	\node[above] at (0,1) {{\small blow-up}};
	\shade[bottom color=gray!50, top color=white] (4.414,3.414-0.707) -- (4.414,1.414-0.707) -- (5.828,0-0.707) -- (7.828,0-0.707) -- (7.828,3.414-0.707);
	\draw (4.414,3.414-0.707) -- (4.414,1.414-0.707) node[draw, circle, inner sep=1pt, fill]{} -- (5.828,0-0.707) node[draw, circle, inner sep=1pt, fill]{} -- (7.828,0-0.707);
	\draw[->] (4.414,2.414-0.707) -- (3.414,2.414-0.707);
	\draw[->] (-0.707+5.828,0.707-0.707) -- (-1.414+5.828,0-0.707);
	\draw[->] (6.828,0-0.707) -- (6.828,-1-0.707);
	\node[above] at (3.414-0.3,2.414-0.707) {$v = (m,n)$};
	\node[below left] at (-1.414+5.828,0-0.707) {$v_0 = (m+m',n+n')$};
	\node[below] at (6.828,-1-0.707) {$v' = (m',n')$};
\end{tikzpicture}
\end{center}
\medskip

We may then compare the action \eqref{Iintro} before and after the blow up. Of course, since the formula is entirely local, this is particularly straightforward.
Suppose the vertex we blow up has chirality $\kappa\in \{\pm 1\}$, and that the chirality of the two associated vertices after the blow up are also $\kappa$ -- 
this is natural if one regards the blow up as a continuous process, where the size of the zero section of $\mc{O}(-1)\rightarrow S^2$ is $r$, with $r\rightarrow 0$ 
being the limit in which one recovers the original geometry with $r=0$. The weights at $p$ are simply
\beq
u_1=(n,-m)\, , \qquad u_2= (-n',m')\, ,
\eeq
while after the blow-up the weights at the two vertices are
\beq
\begin{split}
u_1^{(1)}  & = (n,-m)\, , \qquad  \qquad \ \ \ \ \ \ \  u_2^{(1)} = (-n-n',m+m')\, , \\
u_1^{(2)} & = (n+n',-m-m')\, , \qquad u_2^{(2)} = (-n',m')\, .
\end{split}
\eeq
Again, for generic supersymmetric Killing vector \eqref{eq:xiagain}, 
\eqref{Iintro} gives
\beq
\begin{split}
I_{\mathrm{blow}\, \mathrm{up}}&  = I + \left[\sum_{i=1}^2 -\kappa \frac{(\xi \cdot u_1^{(i)}-\kappa \xi \cdot u_2^{(i)})^2}{4 \xi \cdot u_1^{(i)}\, \xi \cdot u_2^{(i)}}\right]\frac{\pi}{2G_4} - \left[ -\kappa \frac{(\xi \cdot u_1-\kappa \xi \cdot u_2)^2}{4 \xi \cdot u_1\, \xi \cdot u_2}\right]\frac{\pi}{2G_4}\, \\
& = I + \left[\frac{2+3\kappa}{4}\right]\frac{\pi}{2G_4}\, ,
\end{split}
\eeq
where the second line follows from explicit computation and simplifications. 
Remarkably, the action changes by an amount that is \emph{independent} of the choice of supersymmetric Killing vector $\xi=(a_1,a_2)$!
Thus blowing up a vertex with $\kappa=1$ changes the action by $+\tfrac{5}{4}\frac{\pi}{2G_4}$, while blowing up a vertex with $\kappa=-1$ changes 
the action by $-\tfrac{1}{4}\frac{\pi}{2G_4}$. In fact one can verify that the formulas in section \ref{subsec:ToricOp} for $p=1$ are indeed compatible with this result, 
where of course $\mc{O}(-1)\rightarrow S^2$ is the blow up of $\C^2$, where the latter action is given by \eqref{eq:ballaction}. Indeed, the genus $g=0$ 1/4 BPS result \eqref{eq:14BPSAction} for the positive 
branch solution gives $I_+=\left(1 -\frac{1}{4}\right)\frac{\pi}{2G_4}$, which is the value obtained by blowing up Euclidean AdS$_4$, viewed 
as a $\kappa=-1$ solution. As pointed out in \cite{Toldo:2017qsh}, this solution then has lower action than Euclidean AdS$_4$, 
and is thus a more dominant saddle point for $M_3=S^3$ boundary. Interestingly, both saddle points can be seen in the dual field theory calculation, 
although this is currently poorly understood, as pointed out in \cite{Toldo:2017qsh}. Note that if one were able to repeatedly blow up 
a $\kappa=-1$ vertex the action could be made arbitrarily negative, so presumably there is an obstruction to doing this in terms of solving the supergravity 
equations. Indeed, a key assumption above is that a solution actually exists! This is clearly a crucial question, and again we return to discuss this briefly 
in section \ref{sec:Conclusions}. 

\subsection{Supersymmetric black holes}
\label{subsec:BH}

The explicit solutions considered in sections \ref{subsec:14BPS} and \ref{subsec:12BPS} involve a non-trivial fibration over a base surface, which in some cases (for example, choosing $a_1=0$ in \eqref{eq:xiMp}) is a bolt for the supersymmetric Killing vector. However, there are also known solutions where the fibration is trivial and the topology is that of $M=\R^2\times \Sigma_g$ (with a warped metric). One simple class of such solutions that preserve $1/4$ of the supersymmetry can be obtained by Wick rotation of the dyonic static solutions studied in \cite{Romans:1991nq, Brill:1997mf, Caldarelli:1998hg}\footnote{We thank Chiara Toldo for pointing out this class of solutions to us.}
\beq
\label{eq:BHMetric}
\rd s^2 = V(r) \, \rd\tau^2 + \frac{\rd r^2}{V(r)} + r^2 (\rd \vartheta^2 + \sinh^2\vartheta\, \rd\phi^2 ) \, , 
\eeq
with
\beq
V(r) = -1 + \frac{\frac{1}{4}-Q^2}{r^2} + r^2 \, .
\eeq
The two-dimensional metric in round brackets in \eqref{eq:BHMetric}
 is that on $H^2$, and quotienting by discrete subgroups of $SO(1,2)$ we may find Riemann surfaces with any genus $g>1$. The gauge field and its curvature are given by
\beq
\label{eq:BHGaugeField}
A = \frac{Q}{r}\rd \tau + \frac{1}{2}\cosh\vartheta\, \rd\phi \, , \qquad F = \frac{Q}{r^2}\rd\tau \wedge \rd r + \frac{1}{2}\sinh\vartheta\, \rd\vartheta\wedge\rd\phi \, ,
\eeq
so we identify the electric charge with $Q$ and the magnetic charge as
\beq
\label{eq:MagneticChargeBH}
\int_{\Sigma_g}\frac{F}{2\pi} = \frac{1}{4\pi}\int_{\Sigma_g}\vol_{\Sigma_g}=g-1 \, \in \, \Z \, .
\eeq
This last condition ensures that  the gauge field is a connection on an honest $U(1)$ gauge bundle over the spacetime, for any Riemann surface. 

Choosing the vierbein
\beq
\e^1 = r\, \rd\vartheta \, , \quad \e^2 = r\sinh\vartheta\, \rd \phi \, , \quad \e^3 = \sqrt{V}\, \rd\tau \, , \quad \e^4 = \frac{\rd r}{\sqrt{V}} \, ,
\eeq
and using the same $\Gamma$ matrices as in \eqref{eq:GammaMatrices}, we find that the following spinor satisfies the (generalized) Killing spinor equation \eqref{eq:SUSY}
\beq
\label{eq:BHSpinor}
\epsilon = \begin{pmatrix}
0 \\ \sqrt{\frac{2(r-r_-)(r+r_-)}{r}} \chi_{(0)} \\ 0 \\ \ii\sqrt{\frac{2(r-r_+)(r+r_+)}{r}} \chi_{(0)}
\end{pmatrix} \, ,
\eeq
where $\chi_{(0)}$ is a complex constant which we set to $1/2$ to simplify some expressions, and $r_{\pm}$ are the following roots of $V(r)$
\beq
r_{\pm} = \sqrt{\frac{1}{2}\pm Q} \, .
\eeq

\medskip

The solution is regular as long as $V(r)>0$. If we denote the largest root of $V$ by $r_0$ and assume that $Q\neq 0$, then a local analysis near $r_0$ shows that, to avoid conical singularities in the space orthogonal to the bolt, the period of $\tau$ should satisfy the constraint
\beq
\frac{\abs{V'(r_0)}}{2}\Delta\tau = 2\pi \, . 
\eeq
If $Q=0$, instead, $V'(r_0)=0$, so near $r=1/\sqrt{2}$ we introduce the coordinate $\rho= r-1/\sqrt{2}$ and we see that the metric approaches that of $H^2\times \Sigma_g$
\beq
\rd s^2 \sim 4\rho^2 \rd\tau^2 + \frac{\rd\rho^2}{4\rho^2} + \frac{1}{2}\rd s^2_{\Sigma_g} \, ,
\eeq
with an infinite throat as $\rho\rightarrow 0$. 
Therefore, we have a family of solutions depending on a parameter $Q$: as long as $Q\neq 0$, we have a bolt at $r=r_0>1/\sqrt{2}$  for $\partial_{\tau}$ at finite distance, whereas in the case $Q=0$ the ``bolt'' has effectively 
receded to infinite distance. As we shall see, this limit is especially interesting in Lorentzian signature, and we will come back to it after computing the action of these solutions.

Using the spinor \eqref{eq:BHSpinor}, we can immediately see that the supersymmetric Killing vector is $\xi = \partial_{\tau}$. 
We may then compute the functions appearing the three-dimensional reduction of section \ref{sec:4dSUGRA}: in the supersymmetric gauge, where $\j$ and $\nutpot$ satisfy \eqref{eq:jSUSYgauge} and \eqref{eq:SUSYNutPotential2}, we have
\beq
\j = \frac{Q}{r} \, , \qquad \sigma = - \frac{Q}{r^2} + 2Q \, , \qquad \cos\theta = \frac{2Q}{2r^2-1} \, .
\eeq
One can perform the holographic renormalization according to the procedure outlined in subsection \ref{subsec:ConformalBoundary} and see that indeed in this gauge, the contribution to the on-shell action from the conformal infinity is vanishing, and the value is determined by the bolt contribution. Turning instead to our main formula \eqref{Iintro}, 
at the fixed point set $\cos\theta = \sgn(Q)$, so the distinction bolt$_{\pm}$ only depends on the charge $Q$. However, this does not affect the value of the on-shell action, as the chirality label in \eqref{Iintro} only appears in front of the self-intersection number of the bolt, which is zero for the trivial fibration as in this case. Therefore, we conclude from \eqref{Iintro} that the on-shell action for the entire family of solutions \eqref{eq:BHMetric}, \eqref{eq:BHGaugeField} is simply proportional to the Euler characteristic of the bolt
\beq
\label{eq:BHOnShellAction}
I = \frac{\pi}{2G_4}(1-g) \, .
\eeq
Notice that this corresponds to the $p=0$ case of \eqref{eq:14BPSAction}, since the on-shell action only depends on the topological data of the solution. 
Indeed, the action is manifestly independent of $Q$, as it had to be. The fact that \eqref{eq:BHOnShellAction} agrees with the purely magnetically 
charged $Q=0$ black hole action in \cite{Azzurli:2017kxo} then really follows only because the $Q\neq 0$ deformation with a bolt exists: 
as soon as one knows this deformation exists, the action is given by the topological formula \eqref{Iintro}, which is independent of the deformation parameter.
In this way, one can compute the action of solutions with an infinite throat, provided one knows an appropriate deformation of it exists that is closed in the interior. 
Notice that a similar regularization of the action of a solution with an infinite throat, by deforming away from extremality but preserving supersymmetry, was used in various dimensions
in \cite{Cabo-Bizet:2018ehj, Cassani:2019mms}.
Notice also that the relation between gauge field and geometry fixed by \eqref{eq:FluxAndGeometry} is satisfied: the magnetic charge is in this case proportional to the Euler characteristic of the surface, since the normal bundle to the bolt is trivial.

In Lorentzian signature (obtained via $t = \ii \tau$ and $q_e = -\ii Q$), the solutions with $Q\neq 0$ describe dyonic naked singularities, whereas the case $Q=0$ is an extremal magnetically charged black hole with horizon homeomorphic to a Riemann surface. As observed in \cite{Azzurli:2017kxo}, the Bekenstein--Hawking entropy of the black hole is directly reproduced by the Euclidean on-shell action (note that indeed the entropy is positive for genus $g>1$)\footnote{Note the difference with the deformation of the extremal case considered in \cite{Azzurli:2017kxo}: there, the authors consider non-supersymmetric magnetically charged solutions, whereas we have to preserve supersymmetry in order to apply our methods, so we consider dyons. Again, we thank Chiara Toldo for remarks on this solution.}
\beq
I = -S_{\rm BH} \, ,
\eeq
and thus, via the AdS/CFT dictionary, the entropy of the black hole can be connected to the large $N$ limit of the partition function of the boundary SCFT, leading to the relation originally advocated in \cite{Benini:2015eyy}
\beq
S_{\rm BH} = \log Z \, .
\eeq
Moreover, the fact that the formula for the on-shell action factorizes as the action for AdS$_4$, multiplied by the Euler number of the surface, can be interpreted as the manifestation of the boundary universal twist \cite{Azzurli:2017kxo, Benini:2015bwz, Bobev:2017uzs}.

\section{Conclusions}
\label{sec:Conclusions}

Inspired by the analysis in \cite{Gibbons:1979xm}, in this paper we presented a formula \eqref{Iintro} for the on-shell action of supersymmetric asymptotically locally Euclidean AdS  solutions to four-dimensional minimal gauged supergravity in terms of contributions only from the fixed point locus of a canonical supersymmetric Killing vector. The formula is such that it may be evaluated knowing 
only the topology of the four-manifold $M$ and the action generated by the vector field, together with certain signs that are determined by chirality data. We have shown that \eqref{Iintro}
straightforwardly reproduces the actions of explicitly known solutions in the literature, but also that it may be used to simply write down the actions of solutions, \emph{assuming they exist}. 

Recall that the standard holographic dictionary connects the on-shell action to the partition function of a dual field theory defined the boundary manifold $M_3=\partial M$, at least in an appropriate strong coupling limit. 
We may make this more precise in the current setting by 
embedding the construction into string/M-theory, where recall that minimal $\mathcal{N}=2$ gauged supergravity is a consistent truncation of 11-dimensional supergravity on any Sasaki--Einstein 7-manifold $Y_7$ \cite{Gauntlett:2007ma}. Thus, at least locally, any bulk solution on a four-manifold $M$ uplifts to an 11-dimensional solution that is the total space of a fibration of $Y_7\rightarrow M$. 
As discussed in \cite{Martelli:2012sz, Toldo:2017qsh}, globally there are some restrictions on which $Y_7$ may be fibred, depending on the topology of $M$ and the 
spin$^c$ gauge bundle defined by the Abelian gauge field. However, there are by now large classes of Sasaki--Einstein $Y_7$ for which the dual three-dimensional 
superconformal field theories on $M_3=\partial M$ are known explicitly, starting with the seminal work of \cite{Aharony:2008ug} for $Y_7=S^7/\Z_k$. 
These are typically Chern--Simons-matter theories, with the supergravity saddle point limit corresponding to a limit of large rank $N$ of the gauge group. 
As mentioned already, many of the expressions for the on-shell action reviewed in the paper have already been matched to such corresponding field theory calculations, 
showing agreement. However, thanks to \eqref{Iintro} we now have much more general expressions. For example, \eqref{eq:IMp14general} gives the action of solutions 
with topology $M=\mc{O}(-p)\rightarrow S^2$ with a \emph{general} choice of supersymmetric Killing vector, which are continuously connected to the explicit 1/4 BPS 
solution of \cite{Martelli:2012sz} with this topology, which has a particular fixed supersymmetric Killing vector. The formula \eqref{eq:IMp14general} is hence a prediction for the large $N$ limit 
of appropriate classes of Chern--Simons-matter theories on the Lens space $L(p,1)=S^3/\Z_p$, as discussed in section \ref{subsec:ToricOp}. 
The field theory computation required to check this prediction is a generalization of that appearing in \cite{Toldo:2017qsh}, which reproduces the 1/4 BPS 
result \eqref{eq:14BPSAction}.

However, one can go much further. One could start with any Lens space $M_3=L(p,q)$, as discussed in section \ref{subsec:top}, with any choice of 
toric supersymmetric Killing vector \eqref{eq:xiagain}. This defines a rigid supersymmetric three-manifold background \cite{Closset:2012ru},
 and moreover there are now techniques to compute the partition functions of supersymmetric Chern--Simons-matter theories 
on \emph{any} such three-manifold \cite{Closset:2017zgf, Closset:2018ghr} (for rational $a_1/a_2\in\mathbb{Q}$). One can then take the appropriate large $N$ limit to compare to the 
supergravity saddle point result \eqref{Iintro}. 
In \cite{Toldo:2017qsh}, both of the 1/4 BPS bolt$_\pm$ solutions with action \eqref{eq:14BPSAction} were seen quite explicitly in the field theory analysis, despite the fact that 
it is the upper sign branch solution that has the least action. Indeed, for $p=1$ this action is also smaller than the action for the solution with topology $\R^4$, 
which corresponds to yet another large $N$ saddle point solution in field theory. This issue is discussed at length in  \cite{Toldo:2017qsh}. 
It is natural to conjecture that the field theory computation in fact ``sees'' the various supergravity fillings of a given $M_3$, where different supergravity 
fillings can have different topology, but also different chirality data, determining the signs in \eqref{Iintro}.\footnote{These signs were labelled $\kappa_i\in\{\pm1\}$ in sections 
 \ref{subsec:ToricOp} and \ref{subsec:top}.} Since it is known how to perform these field theory computations, this is perhaps the most immediate 
and interesting direction to pursue given the results of this work. For example, an immediate problem is to reproduce the $L(3,2)$ filling action in 
\eqref{eq:I32action}. Moreover, one might ask if there is in a precise sense a gravity dual of the Seifert fibering operators of \cite{Closset:2018ghr}, that 
change the topology of the Seifert three-manifold boundary $M_3$, and correspondingly lead to a change in the filling four-manifold and action \eqref{Iintro}. 

In the absence of explicit solutions, application of \eqref{Iintro} also requires us to assume such a supergravity solution actually exists, which leads naturally to the question 
of existence and uniqueness of solutions. We note this is similar to the situation in \cite{Martelli:2005tp}, where the volumes of (toric) Sasaki--Einstein manifolds 
could be computed explicitly, again assuming that the Einstein equation actually has a solution. In this setting, reference \cite{Futaki:2006cc} 
subsequently proved that such solutions do indeed always exist. This is in general a problem in geometric analysis. However, we also note that
one might more simply address this existence problem for self-dual solutions, as briefly summarized in section \eqref{subsec:SD}, since the PDE \eqref{eq:Toda}
in this case is integrable. Indeed, formally infinite families of  solutions to this equation may be written down, where $M$ has a $T^2$ isometry, as summarized in section 5.4 of \cite{Farquet:2014kma}, following 
\cite{Calderbank:2001uz}. Given that local solutions satisfying the supergravity equations are trivial to construct within this ansatz, this is 
likely the best place to begin to answer these (global) existence questions. 

Of course, in the context of holographic approaches to quantum gravity, it would also be very interesting to consider the subleading corrections in the rank of the gauge group of the boundary theory, as done for instance in \cite{Liu:2017vbl, Liu:2017vll, Liu:2018bac,  Gang:2019uay}. In particular, we notice that in the latter reference the corrections have been computed for the minimal gauged supergravity considered here, for the black hole solution \eqref{eq:BHMetric}, even though there it was seen in the context of a reduction from 7-dimensional supergravity with a view to holography and the 3d/3d correspondence.

In a different direction, a natural generalization of this work would consider different, less simple, supergravity theories. This is particularly motivated by holographic computations of the entropy of black holes. For instance, in agreement with \cite{Azzurli:2017kxo}, we found in section \ref{subsec:BH} a relation between the on-shell action and the Bekenstein--Hawking entropy of the Lorentzian black hole. {\it A priori}, the standard gauge/gravity dictionary relates the bulk on-shell action to the boundary partition function. However, in models where gravity is coupled to vector multiplets (namely the STU model), a relation, involving a Legendre transform, between the black hole entropy and the supersymmetric partition function computed via localization has been originally advocated in \cite{Benini:2015bwz}, and then shown to be a consequence of the BPS relation \cite{Halmagyi:2017hmw, Cabo-Bizet:2017xdr}. It would be interesting to see if there is a generalization of the structure underlying supersymmetric solutions that we have found here, see e.g. \cite{Benini:2015bwz, Hosseini:2016tor, Hosseini:2016ume, Benini:2016hjo, Benini:2016rke, Cabo-Bizet:2017jsl, Azzurli:2017kxo, Hosseini:2017fjo, Benini:2017oxt, Bobev:2018uxk, Hristov:2018spe}. 

Finally, as already remarked in the introduction, supersymmetric localization for field theories on curved backgrounds has allowed for spectacular improvements in our understanding of quantum field theories at strong coupling, and the duality relations that appear in that regime. 
Here we have presented a formula \eqref{Iintro} that suggests that a localization similar in spirit already happens in classical supergravity. It is interesting to speculate whether there is a more precise connection between the boundary and bulk computations, and indeed whether the formula \eqref{Iintro} can be understood directly from a large $N$ field theory computation 
on the boundary three-manifold $M_3=\partial M$. 

\section*{Acknowledgments}
We are grateful to Gary Gibbons, Dario Martelli, Harvey Reall and Chiara Toldo for helpful conversations. The work of PBG has been supported by the STFC consolidated grant ST/P000681/1.

\bibliographystyle{./OnShellActionFiles/JHEP}
\bibliography{./OnShellActionFiles/Bib_OnShellAction}

\providecommand{\href}[2]{#2}\begingroup\raggedright\begin{thebibliography}{10}

\bibitem{Pestun:2016zxk}
V.~Pestun et~al., \emph{{Localization techniques in quantum field theories}},
  \href{http://dx.doi.org/10.1088/1751-8121/aa63c1}{\emph{J. Phys.} {\bf A50}
  (2017) 440301}, [\href{http://arxiv.org/abs/1608.02952}{{\tt 1608.02952}}].

\bibitem{Dabholkar:2014wpa}
A.~Dabholkar, N.~Drukker and J.~Gomes, \emph{{Localization in supergravity and
  quantum $AdS_4/CFT_3$ holography}},
  \href{http://dx.doi.org/10.1007/JHEP10(2014)090}{\emph{JHEP} {\bf 10} (2014)
  90}, [\href{http://arxiv.org/abs/1406.0505}{{\tt 1406.0505}}].

\bibitem{Hristov:2018lod}
K.~Hristov, I.~Lodato and V.~Reys, \emph{{On the quantum entropy function in 4d
  gauged supergravity}},
  \href{http://dx.doi.org/10.1007/JHEP07(2018)072}{\emph{JHEP} {\bf 07} (2018)
  072}, [\href{http://arxiv.org/abs/1803.05920}{{\tt 1803.05920}}].

\bibitem{deWit:2018dix}
B.~de~Wit, S.~Murthy and V.~Reys, \emph{{BRST quantization and equivariant
  cohomology: localization with asymptotic boundaries}},
  \href{http://dx.doi.org/10.1007/JHEP09(2018)084}{\emph{JHEP} {\bf 09} (2018)
  084}, [\href{http://arxiv.org/abs/1806.03690}{{\tt 1806.03690}}].

\bibitem{Jeon:2018kec}
I.~Jeon and S.~Murthy, \emph{{Twisting and localization in supergravity:
  equivariant cohomology of BPS black holes}},
  \href{http://dx.doi.org/10.1007/JHEP03(2019)140}{\emph{JHEP} {\bf 03} (2019)
  140}, [\href{http://arxiv.org/abs/1806.04479}{{\tt 1806.04479}}].

\bibitem{Nekrasov:2002qd}
N.~A. Nekrasov, \emph{{Seiberg-Witten prepotential from instanton counting}},
  \href{http://dx.doi.org/10.4310/ATMP.2003.v7.n5.a4}{\emph{Adv. Theor. Math.
  Phys.} {\bf 7} (2003) 831--864},
  [\href{http://arxiv.org/abs/hep-th/0206161}{{\tt hep-th/0206161}}].

\bibitem{Freedman:1976aw}
D.~Z. Freedman and A.~K. Das, \emph{{Gauge Internal Symmetry in Extended
  Supergravity}},
  \href{http://dx.doi.org/10.1016/0550-3213(77)90041-4}{\emph{Nucl. Phys.} {\bf
  B120} (1977) 221--230}.

\bibitem{Gauntlett:2007ma}
J.~P. Gauntlett and O.~Varela, \emph{{Consistent Kaluza-Klein reductions for
  general supersymmetric AdS solutions}},
  \href{http://dx.doi.org/10.1103/PhysRevD.76.126007}{\emph{Phys. Rev.} {\bf
  D76} (2007) 126007}, [\href{http://arxiv.org/abs/0707.2315}{{\tt
  0707.2315}}].

\bibitem{Martelli:2011fu}
D.~Martelli, A.~Passias and J.~Sparks, \emph{{The gravity dual of
  supersymmetric gauge theories on a squashed three-sphere}},
  \href{http://dx.doi.org/10.1016/j.nuclphysb.2012.07.019}{\emph{Nucl. Phys.}
  {\bf B864} (2012) 840--868}, [\href{http://arxiv.org/abs/1110.6400}{{\tt
  1110.6400}}].

\bibitem{Martelli:2011fw}
D.~Martelli and J.~Sparks, \emph{{The gravity dual of supersymmetric gauge
  theories on a biaxially squashed three-sphere}},
  \href{http://dx.doi.org/10.1016/j.nuclphysb.2012.08.015}{\emph{Nucl. Phys.}
  {\bf B866} (2013) 72--85}, [\href{http://arxiv.org/abs/1111.6930}{{\tt
  1111.6930}}].

\bibitem{Martelli:2012sz}
D.~Martelli, A.~Passias and J.~Sparks, \emph{{The supersymmetric NUTs and bolts
  of holography}},
  \href{http://dx.doi.org/10.1016/j.nuclphysb.2013.04.026}{\emph{Nucl. Phys.}
  {\bf B876} (2013) 810--870}, [\href{http://arxiv.org/abs/1212.4618}{{\tt
  1212.4618}}].

\bibitem{Martelli:2013aqa}
D.~Martelli and A.~Passias, \emph{{The gravity dual of supersymmetric gauge
  theories on a two-parameter deformed three-sphere}},
  \href{http://dx.doi.org/10.1016/j.nuclphysb.2013.09.012}{\emph{Nucl. Phys.}
  {\bf B877} (2013) 51--72}, [\href{http://arxiv.org/abs/1306.3893}{{\tt
  1306.3893}}].

\bibitem{Farquet:2014kma}
D.~Farquet, J.~Lorenzen, D.~Martelli and J.~Sparks, \emph{{Gravity duals of
  supersymmetric gauge theories on three-manifolds}},
  \href{http://dx.doi.org/10.1007/JHEP08(2016)080}{\emph{JHEP} {\bf 08} (2016)
  080}, [\href{http://arxiv.org/abs/1404.0268}{{\tt 1404.0268}}].

\bibitem{Azzurli:2017kxo}
F.~Azzurli, N.~Bobev, P.~M. Crichigno, V.~S. Min and A.~Zaffaroni, \emph{{A
  universal counting of black hole microstates in AdS$_{4}$}},
  \href{http://dx.doi.org/10.1007/JHEP02(2018)054}{\emph{JHEP} {\bf 02} (2018)
  054}, [\href{http://arxiv.org/abs/1707.04257}{{\tt 1707.04257}}].

\bibitem{Toldo:2017qsh}
C.~Toldo and B.~Willett, \emph{{Partition functions on 3d circle bundles and
  their gravity duals}},
  \href{http://dx.doi.org/10.1007/JHEP05(2018)116}{\emph{JHEP} {\bf 05} (2018)
  116}, [\href{http://arxiv.org/abs/1712.08861}{{\tt 1712.08861}}].

\bibitem{Closset:2012ru}
C.~Closset, T.~T. Dumitrescu, G.~Festuccia and Z.~Komargodski,
  \emph{{Supersymmetric Field Theories on Three-Manifolds}},
  \href{http://dx.doi.org/10.1007/JHEP05(2013)017}{\emph{JHEP} {\bf 05} (2013)
  017}, [\href{http://arxiv.org/abs/1212.3388}{{\tt 1212.3388}}].

\bibitem{Emparan:1999pm}
R.~Emparan, C.~V. Johnson and R.~C. Myers, \emph{{Surface terms as counterterms
  in the AdS / CFT correspondence}},
  \href{http://dx.doi.org/10.1103/PhysRevD.60.104001}{\emph{Phys. Rev.} {\bf
  D60} (1999) 104001}, [\href{http://arxiv.org/abs/hep-th/9903238}{{\tt
  hep-th/9903238}}].

\bibitem{Skenderis:2002wp}
K.~Skenderis, \emph{{Lecture notes on holographic renormalization}},
  \href{http://dx.doi.org/10.1088/0264-9381/19/22/306}{\emph{Class. Quant.
  Grav.} {\bf 19} (2002) 5849--5876},
  [\href{http://arxiv.org/abs/hep-th/0209067}{{\tt hep-th/0209067}}].

\bibitem{Gibbons:1979xm}
G.~W. Gibbons and S.~W. Hawking, \emph{{Classification of Gravitational
  Instanton Symmetries}},
  \href{http://dx.doi.org/10.1007/BF01197189}{\emph{Commun. Math. Phys.} {\bf
  66} (1979) 291--310}.

\bibitem{Dunajski:2010uv}
M.~Dunajski, J.~B. Gutowski, W.~A. Sabra and P.~Tod, \emph{{Cosmological
  Einstein-Maxwell Instantons and Euclidean Supersymmetry: Beyond
  Self-Duality}}, \href{http://dx.doi.org/10.1007/JHEP03(2011)131}{\emph{JHEP}
  {\bf 03} (2011) 131}, [\href{http://arxiv.org/abs/1012.1326}{{\tt
  1012.1326}}].

\bibitem{Genolini:2016ecx}
P.~Benetti~Genolini, D.~Cassani, D.~Martelli and J.~Sparks, \emph{{Holographic
  renormalization and supersymmetry}},
  \href{http://dx.doi.org/10.1007/JHEP02(2017)132}{\emph{JHEP} {\bf 02} (2017)
  132}, [\href{http://arxiv.org/abs/1612.06761}{{\tt 1612.06761}}].

\bibitem{Dunajski:2010zp}
M.~Dunajski, J.~Gutowski, W.~Sabra and P.~Tod, \emph{{Cosmological
  Einstein-Maxwell Instantons and Euclidean Supersymmetry: Anti-Self-Dual
  Solutions}},
  \href{http://dx.doi.org/10.1088/0264-9381/28/2/025007}{\emph{Class. Quant.
  Grav.} {\bf 28} (2011) 025007}, [\href{http://arxiv.org/abs/1006.5149}{{\tt
  1006.5149}}].

\bibitem{CEJM}
A.~Chamblin, R.~Emparan, C.~V. Johnson and R.~C. Myers, \emph{{Large N phases,
  gravitational instantons and the nuts and bolts of AdS holography}},
  {\emph{Phys. Rev.} {\bf D59} (1999) 064010},
  [\href{http://arxiv.org/abs/9808177}{{\tt 9808177}}].

\bibitem{OrlikRaymond}
P.~Orlik and F.~Raymond, \emph{Actions of the torus on 4-manifolds. i},
  {\emph{Transactions of the American Mathematical Society} {\bf 152} (1970)
  531--559}.

\bibitem{Calderbank:2002gy}
D.~M.~J. Calderbank and M.~A. Singer, \emph{{Einstein metrics and complex
  singularities}},
  \href{http://dx.doi.org/10.1007/s00222-003-0344-1}{\emph{Invent. Math.} {\bf
  156} (2004) 405}, [\href{http://arxiv.org/abs/math/0206229}{{\tt
  math/0206229}}].

\bibitem{Romans:1991nq}
L.~J. Romans, \emph{{Supersymmetric, cold and lukewarm black holes in
  cosmological Einstein-Maxwell theory}},
  \href{http://dx.doi.org/10.1016/0550-3213(92)90684-4}{\emph{Nucl. Phys.} {\bf
  B383} (1992) 395--415}, [\href{http://arxiv.org/abs/hep-th/9203018}{{\tt
  hep-th/9203018}}].

\bibitem{Brill:1997mf}
D.~R. Brill, J.~Louko and P.~Peldan, \emph{{Thermodynamics of (3+1)-dimensional
  black holes with toroidal or higher genus horizons}},
  \href{http://dx.doi.org/10.1103/PhysRevD.56.3600}{\emph{Phys. Rev.} {\bf D56}
  (1997) 3600--3610}, [\href{http://arxiv.org/abs/gr-qc/9705012}{{\tt
  gr-qc/9705012}}].

\bibitem{Caldarelli:1998hg}
M.~M. Caldarelli and D.~Klemm, \emph{{Supersymmetry of Anti-de Sitter black
  holes}}, \href{http://dx.doi.org/10.1016/S0550-3213(98)00846-3}{\emph{Nucl.
  Phys.} {\bf B545} (1999) 434--460},
  [\href{http://arxiv.org/abs/hep-th/9808097}{{\tt hep-th/9808097}}].

\bibitem{Cabo-Bizet:2018ehj}
A.~Cabo-Bizet, D.~Cassani, D.~Martelli and S.~Murthy, \emph{{Microscopic origin
  of the Bekenstein-Hawking entropy of supersymmetric AdS$_{\bf 5}$ black
  holes}},  \href{http://arxiv.org/abs/1810.11442}{{\tt 1810.11442}}.

\bibitem{Cassani:2019mms}
D.~Cassani and L.~Papini, \emph{{The BPS limit of rotating AdS black hole
  thermodynamics}},  \href{http://arxiv.org/abs/1906.10148}{{\tt 1906.10148}}.

\bibitem{Benini:2015eyy}
F.~Benini, K.~Hristov and A.~Zaffaroni, \emph{{Black hole microstates in
  AdS$_{4}$ from supersymmetric localization}},
  \href{http://dx.doi.org/10.1007/JHEP05(2016)054}{\emph{JHEP} {\bf 05} (2016)
  054}, [\href{http://arxiv.org/abs/1511.04085}{{\tt 1511.04085}}].

\bibitem{Benini:2015bwz}
F.~Benini, N.~Bobev and P.~M. Crichigno, \emph{{Two-dimensional SCFTs from
  D3-branes}}, \href{http://dx.doi.org/10.1007/JHEP07(2016)020}{\emph{JHEP}
  {\bf 07} (2016) 020}, [\href{http://arxiv.org/abs/1511.09462}{{\tt
  1511.09462}}].

\bibitem{Bobev:2017uzs}
N.~Bobev and P.~M. Crichigno, \emph{{Universal RG Flows Across Dimensions and
  Holography}}, \href{http://dx.doi.org/10.1007/JHEP12(2017)065}{\emph{JHEP}
  {\bf 12} (2017) 065}, [\href{http://arxiv.org/abs/1708.05052}{{\tt
  1708.05052}}].

\bibitem{Aharony:2008ug}
O.~Aharony, O.~Bergman, D.~L. Jafferis and J.~Maldacena, \emph{{N=6
  superconformal Chern-Simons-matter theories, M2-branes and their gravity
  duals}}, \href{http://dx.doi.org/10.1088/1126-6708/2008/10/091}{\emph{JHEP}
  {\bf 10} (2008) 091}, [\href{http://arxiv.org/abs/0806.1218}{{\tt
  0806.1218}}].

\bibitem{Closset:2017zgf}
C.~Closset, H.~Kim and B.~Willett, \emph{{Supersymmetric partition functions
  and the three-dimensional A-twist}},
  \href{http://dx.doi.org/10.1007/JHEP03(2017)074}{\emph{JHEP} {\bf 03} (2017)
  074}, [\href{http://arxiv.org/abs/1701.03171}{{\tt 1701.03171}}].

\bibitem{Closset:2018ghr}
C.~Closset, H.~Kim and B.~Willett, \emph{{Seifert fibering operators in 3d
  $\mathcal{N}=2$ theories}},
  \href{http://dx.doi.org/10.1007/JHEP11(2018)004}{\emph{JHEP} {\bf 11} (2018)
  004}, [\href{http://arxiv.org/abs/1807.02328}{{\tt 1807.02328}}].

\bibitem{Martelli:2005tp}
D.~Martelli, J.~Sparks and S.-T. Yau, \emph{{The Geometric dual of
  a-maximisation for Toric Sasaki-Einstein manifolds}},
  \href{http://dx.doi.org/10.1007/s00220-006-0087-0}{\emph{Commun. Math. Phys.}
  {\bf 268} (2006) 39--65}, [\href{http://arxiv.org/abs/hep-th/0503183}{{\tt
  hep-th/0503183}}].

\bibitem{Futaki:2006cc}
A.~Futaki, H.~Ono and G.~Wang, \emph{{Transverse Kahler geometry of Sasaki
  manifolds and toric Sasaki-Einstein manifolds}}, {\emph{J. Diff. Geom.} {\bf
  83} (2009) 585--636}, [\href{http://arxiv.org/abs/math/0607586}{{\tt
  math/0607586}}].

\bibitem{Calderbank:2001uz}
D.~M.~J. Calderbank and H.~Pedersen, \emph{{Selfdual Einstein metrics with
  torus symmetry}}, {\emph{J. Diff. Geom.} {\bf 60} (2002) 485--521},
  [\href{http://arxiv.org/abs/math/0105263}{{\tt math/0105263}}].

\bibitem{Liu:2017vbl}
J.~T. Liu, L.~A. Pando~Zayas, V.~Rathee and W.~Zhao, \emph{{One-Loop Test of
  Quantum Black Holes in anti–de Sitter Space}},
  \href{http://dx.doi.org/10.1103/PhysRevLett.120.221602}{\emph{Phys. Rev.
  Lett.} {\bf 120} (2018) 221602}, [\href{http://arxiv.org/abs/1711.01076}{{\tt
  1711.01076}}].

\bibitem{Liu:2017vll}
J.~T. Liu, L.~A. Pando~Zayas, V.~Rathee and W.~Zhao, \emph{{Toward Microstate
  Counting Beyond Large N in Localization and the Dual One-loop Quantum
  Supergravity}}, \href{http://dx.doi.org/10.1007/JHEP01(2018)026}{\emph{JHEP}
  {\bf 01} (2018) 026}, [\href{http://arxiv.org/abs/1707.04197}{{\tt
  1707.04197}}].

\bibitem{Liu:2018bac}
J.~T. Liu, L.~A. Pando~Zayas and S.~Zhou, \emph{{Subleading Microstate Counting
  in the Dual to Massive Type IIA}},
  \href{http://arxiv.org/abs/1808.10445}{{\tt 1808.10445}}.

\bibitem{Gang:2019uay}
D.~Gang, N.~Kim and L.~A. Pando~Zayas, \emph{{Precision Microstate Counting for
  the Entropy of Wrapped M5-branes}},
  \href{http://arxiv.org/abs/1905.01559}{{\tt 1905.01559}}.

\bibitem{Halmagyi:2017hmw}
N.~Halmagyi and S.~Lal, \emph{{On the on-shell: the action of AdS$_{4}$ black
  holes}}, \href{http://dx.doi.org/10.1007/JHEP03(2018)146}{\emph{JHEP} {\bf
  03} (2018) 146}, [\href{http://arxiv.org/abs/1710.09580}{{\tt 1710.09580}}].

\bibitem{Cabo-Bizet:2017xdr}
A.~Cabo-Bizet, U.~Kol, L.~A. Pando~Zayas, I.~Papadimitriou and V.~Rathee,
  \emph{{Entropy functional and the holographic attractor mechanism}},
  \href{http://dx.doi.org/10.1007/JHEP05(2018)155}{\emph{JHEP} {\bf 05} (2018)
  155}, [\href{http://arxiv.org/abs/1712.01849}{{\tt 1712.01849}}].

\bibitem{Hosseini:2016tor}
S.~M. Hosseini and A.~Zaffaroni, \emph{{Large $N$ matrix models for 3d ${\cal
  N}=2$ theories: twisted index, free energy and black holes}},
  \href{http://dx.doi.org/10.1007/JHEP08(2016)064}{\emph{JHEP} {\bf 08} (2016)
  064}, [\href{http://arxiv.org/abs/1604.03122}{{\tt 1604.03122}}].

\bibitem{Hosseini:2016ume}
S.~M. Hosseini and N.~Mekareeya, \emph{{Large $N$ topologically twisted index:
  necklace quivers, dualities, and Sasaki-Einstein spaces}},
  \href{http://dx.doi.org/10.1007/JHEP08(2016)089}{\emph{JHEP} {\bf 08} (2016)
  089}, [\href{http://arxiv.org/abs/1604.03397}{{\tt 1604.03397}}].

\bibitem{Benini:2016hjo}
F.~Benini and A.~Zaffaroni, \emph{{Supersymmetric partition functions on
  Riemann surfaces}}, {\emph{Proc. Symp. Pure Math.} {\bf 96} (2017) 13--46},
  [\href{http://arxiv.org/abs/1605.06120}{{\tt 1605.06120}}].

\bibitem{Benini:2016rke}
F.~Benini, K.~Hristov and A.~Zaffaroni, \emph{{Exact microstate counting for
  dyonic black holes in AdS4}},
  \href{http://dx.doi.org/10.1016/j.physletb.2017.05.076}{\emph{Phys. Lett.}
  {\bf B771} (2017) 462--466}, [\href{http://arxiv.org/abs/1608.07294}{{\tt
  1608.07294}}].

\bibitem{Cabo-Bizet:2017jsl}
A.~Cabo-Bizet, V.~I. Giraldo-Rivera and L.~A. Pando~Zayas, \emph{{Microstate
  counting of AdS$_{4}$ hyperbolic black hole entropy via the topologically
  twisted index}}, \href{http://dx.doi.org/10.1007/JHEP08(2017)023}{\emph{JHEP}
  {\bf 08} (2017) 023}, [\href{http://arxiv.org/abs/1701.07893}{{\tt
  1701.07893}}].

\bibitem{Hosseini:2017fjo}
S.~M. Hosseini, K.~Hristov and A.~Passias, \emph{{Holographic microstate
  counting for AdS$_{4}$ black holes in massive IIA supergravity}},
  \href{http://dx.doi.org/10.1007/JHEP10(2017)190}{\emph{JHEP} {\bf 10} (2017)
  190}, [\href{http://arxiv.org/abs/1707.06884}{{\tt 1707.06884}}].

\bibitem{Benini:2017oxt}
F.~Benini, H.~Khachatryan and P.~Milan, \emph{{Black hole entropy in massive
  Type IIA}}, \href{http://dx.doi.org/10.1088/1361-6382/aa9f5b}{\emph{Class.
  Quant. Grav.} {\bf 35} (2018) 035004},
  [\href{http://arxiv.org/abs/1707.06886}{{\tt 1707.06886}}].

\bibitem{Bobev:2018uxk}
N.~Bobev, V.~S. Min and K.~Pilch, \emph{{Mass-deformed ABJM and black holes in
  AdS$_{4}$}}, \href{http://dx.doi.org/10.1007/JHEP03(2018)050}{\emph{JHEP}
  {\bf 03} (2018) 050}, [\href{http://arxiv.org/abs/1801.03135}{{\tt
  1801.03135}}].

\bibitem{Hristov:2018spe}
K.~Hristov, S.~Katmadas and C.~Toldo, \emph{{Rotating attractors and BPS black
  holes in $AdS_4$}},
  \href{http://dx.doi.org/10.1007/JHEP01(2019)199}{\emph{JHEP} {\bf 01} (2019)
  199}, [\href{http://arxiv.org/abs/1811.00292}{{\tt 1811.00292}}].

\end{thebibliography}\endgroup

\end{document}